\documentclass[11pt]{article}

  \usepackage[a4paper,hmarginratio=1:1,vmarginratio=2:3,totalwidth=16cm,
  totalheight=20.6cm]{geometry}  

\usepackage{latexsym}

\usepackage{epsfig}

\newcommand{\cE}{{\cal E}}

\newcommand{\cG}{{\cal G}}

\newcommand{\cJ}{{\cal J}}

\newcommand{\cL}{{\cal L}}

\newcommand{\cO}{{\cal O}}

\newcommand{\bZ}{{\bf Z}}

\newcommand{\Li}{\mbox{Li}}

\newcommand{\ra}{\rightarrow}
\newcommand{\be}{\begin{equation}}
\newcommand{\ee}{\end{equation}}
\newcommand{\bea}{\begin{eqnarray}}
\newcommand{\eea}{\end{eqnarray}}

\DeclareMathSymbol{\mg}{\mathrel}{symbols}{"1D}

\newcounter{oldcounter} 
\addtocounter{equation}{1}
\begin{document} 
\begin{flushright}  
{DAMTP-2005-34}\\
{DESY-05-073}
\end{flushright}  
\vskip 3 cm
\begin{center} 
{\Large 
{\bf  Higher Derivative Operators from Transmission of\\
\vspace{0.2cm}
 Supersymmetry Breaking on $S_1/Z_2$. 
\\}} 
\vspace{1.cm} 
{\bf D.M. Ghilencea$^a$} and {\bf Hyun Min Lee$^b$}\\
\vspace{0.8cm} 
{\it $^a $D.A.M.T.P., Centre for Mathematical Sciences, 
University of Cambridge, \\
Wilberforce Road, Cambridge CB3 OWA, United Kingdom}\\
\vspace{0.4cm}
{\it $^b $Deutsches Elektronen-Synchrotron DESY, \\
D-22603 Hamburg, Germany}\\
\end{center} 
\vspace{0.5cm}
\begin{abstract}
\noindent
We discuss the role that higher derivative operators
play in field theory orbifold compactifications on $S_1/Z_2$
with local and non-local Scherk-Schwarz breaking of supersymmetry.  
Integrating out the bulk fields generates brane-localised  higher derivative 
counterterms to the mass of the brane (or zero-mode of the bulk)
scalar field, identified with the Higgs field  in many
realistic models.  Both Yukawa and gauge interactions are considered  and
the one-loop results found can be used to study the ``running'' of
the scalar field mass with respect to the momentum scale in 5D orbifolds. 
In particular this allows the study of  the behaviour of the 
 mass  under UV scaling of the
momentum. The relation between supersymmetry breaking and
the presence  of higher derivative counterterms to the mass of the
scalar field is investigated. 
This shows that, regardless of the breaking mechanism,
(initial) supersymmetry cannot, in general,  prevent the emergence
of such operators. Some implications for phenomenology of the higher 
derivative operators are also presented.

\end{abstract}
\newpage


\section{Introduction}
The physics of extra dimensions is receiving a
strong research interest in  the context of effective field
theory approaches to compactification, sometimes referred to as
``field theory orbifolds''. While string theory may provide the ultimate 
description of compactification, field theory orbifolds
can  also consistently (re)address  interesting issues  
such as supersymmetry breaking, the hierarchy problem, radiative 
corrections or their experimental  signatures.
In this work we investigate the implications  of supersymmetry
breaking in 5D field theory orbifolds for the
one-loop corrections to the mass of the  4D scalar fields.

An aspect that is somewhat overlooked in recent 
studies of radiative  corrections in gauge theories on
field theory  orbifolds (see for example 
\cite{Antoniadis}-\cite{Delgado:2001xr}),
is the role  of higher derivative 
operators\footnote{For  works on this topic  see for example 
\cite{Ghilencea:2003xj}-\cite{Ghilencea:2004sq}.}.
In general it is thought that such operators, being higher dimensional, 
are suppressed if the scale where  they become relevant is high enough.  
However, from a 4D point of view the most natural such  scale
 is the compactification scale $1/R$ and, if this 
is low (TeV scale),  such operators can have a 
significant effect even at low energies.
For generality we shall consider the role of such operators in 
orbifolds with a radius $R$ of an arbitrary  value (large or small).

The models that we investigate are  5D N=1 supersymmetric,
 compactified on $S_1/Z_2$ and the interactions are 
 localised 
superpotential (in the extra dimension) and gauge interactions. These 
are standard interactions  in any higher dimensional 
theory which aims to recover after compactification and at low
energies, the  Standard Model (SM) or its supersymmetric versions. 
Such interactions  induce loop corrections  
to the mass of the zero mode scalar field (if this is a bulk 
field) or to that of a brane scalar field. It is important to note 
that this field is regarded in many models
as the SM Higgs field \cite{Antoniadis}-\cite{DiClemente:2002qa}. 
Therefore such loop corrections are important for the hierarchy
problem, electroweak symmetry breaking or running of the mass
of the scalar field.
 
Recent studies  of loop corrections  to the mass of the 
4D scalar field showed that radiative corrections are of type 
(see for example \cite{Antoniadis,Arkani-Hamed:2001mi})
\begin{eqnarray}\label{one-loop_c1}
m_{\phi_H}^2\sim \eta\,\frac{\alpha^2}{R^2},
\end{eqnarray}
where $\eta=-1\,(+1)$ when $\alpha$ is the Yukawa (gauge) 
coupling.
As a result one can have radiative electroweak symmetry
breaking\footnote{assuming a vanishing mass at the tree level.}
triggered by towers of Kaluza-Klein states 
of bulk fields which couple, via the aforementioned interactions, 
to the (zero-mode or brane) scalar field associated with the Higgs
field. Note that in the absence of a rigorous
mechanism  to fix $1/R$  to  low values 
(TeV or so) such results do not provide a solution to the
hierarchy problem even at one-loop. But they 
open new ways to address these problems and we think 
this is worth further investigation.  

As we shall see shortly, it turns out that the non-renormalisable
character of these theories becomes apparent earlier than one 
may like, but with potentially interesting consequences.
That means  that results like (\ref{one-loop_c1}) are in general 
altered  by the effects of higher derivative  operators, in some cases
already at one loop.  Their effect  depends on the size of the  
radius $R$, but as already mentioned, here we keep $R$ arbitrary.  
The result of our one-loop Yukawa corrections shows that 
\begin{eqnarray}\label{one-loop_c2}
m_{\phi_H}^2(q^2)=m_{\phi_H}^2(0)+\xi\, q^4
\,R^2+\frac{1}{R^2} \,\cO(q^2 R^2),\qquad m_{\phi_H}^2(0)\sim
-\frac{\alpha^2}{R^2},
\end{eqnarray}
where $\xi$ is an (unknown) coefficient of the higher derivative
operator.

Eq.(\ref{one-loop_c2})  shows that the scalar mass depends on both $R$ and
$1/R$ which\footnote{Such dependence of loop corrections 
on both $R$ and $1/R$   is familiar in one-loop string 
corrections (T duality) such as those to gauge couplings
\cite{DKL}.}
 we find rather interesting.  Further, if $R$ is somehow fixed
to a large value (inverse TeV scale)  to explain a low mass for the
scalar without large fine tuning,   the second term in the first equation 
becomes more important. Given the unknown coefficient $\xi$ of the higher
derivative operators, this can affect  the predictive 
power of the models. Further difficulties at consistency level may
arise for special values of this parameter, as we discuss in
Section~\ref{section4}.  Conversely, if $R$ is very small ($q^2$ fixed), 
the role of higher derivative operators is suppressed, but 
then the first term 
re-introduces the quadratic mass scale (hierarchy) problem 
at one-loop,  familiar from the Standard Model. While this is 
the general picture, a detail analysis should consider the $\cO(q^2
R^2)$  terms in (\ref{one-loop_c2}) whose expression can be computed
in general cases using our technical results. 
 Finally, our  calculation can be used to  re-address previous studies
 of the radiative electroweak symmetry breaking induced by towers 
of Kaluza-Klein modes,  and to investigate the one-loop running of 
the  scalar mass and its ultraviolet behaviour under the 
UV scaling of the   momenta, $q^2\ra \sigma q^2$.  

We  investigate whether the presence of higher derivative operators 
and the above considerations depend on the  way 5D N=1 supersymmetry is
broken. This is the main purpose of the paper.
We consider both local
and non-local supersymmetry breaking, 
transmitted to the visible sector via radiative
corrections. The non-local breaking 
includes discrete and  continuous Scherk-Schwarz twisted boundary 
conditions for the bulk fields. The plan of the paper is as follows:
in Section~\ref{section2} we discuss the orbifold compactifications
and supersymmetry breaking by local  $F$ terms of a brane field at a
hidden brane  (Section~\ref{section2.1}) and 
non-local breaking (Section~\ref{section2.2}) by Scherk-Schwarz 
boundary conditions.
We  show that higher derivative operators are generated as counterterms
for Yukawa but not for gauge interactions. 
It turns out that the presence of these counterterms is
rather independent of the above ways of supersymmetry 
breaking and this is discussed in 
Section~\ref{section3}. Some phenomenological implications 
of the higher derivative operators that are found 
  are discussed in Section~\ref{section4}.  
Appendix~\ref{appendixA} contains formulae relevant
for the  study  of the running of the scalar field mass (with the
momentum scale), which may also be used in other 
applications. Appendix~\ref{appendixB} 
outlines a dimensional analysis needed in Section~\ref{section4}.

\section{Higher derivative counterterms on $S_1/Z_2$.}\label{section2}

On the $S_1/Z_2$ orbifold, one can consider 
vector supermultiplets and hypermultiplets.  
The former may  be described in a 4D language as
made of vector superfield  $V(\lambda_1,A_\mu)$ and an adjoint chiral
superfield $\Sigma((\sigma+i A_5)/\sqrt{2}, \lambda_2)$, where
$\lambda_{1,2}$ are Weyl fermions, $\sigma$ a real scalar and $A_\mu,
A_5$ the 5D gauge field. The hypermultiplet contains two
 chiral superfields $\Phi(\phi,\psi)$ and $\Phi^c(\phi^c, \psi^c)$
with opposite quantum numbers, and where $\phi, \phi^c$ are complex
scalars and $\psi,\psi^c$ are Weyl fermions.  The orbifold
conditions considered are such as the gauge field $A_\mu$ has even parity
(has a massless zero mode) to respect the 4D gauge invariance. 
We consider the following parity assignments
\begin{eqnarray}\label{orbifold}
\Phi(x, - y)=\Phi(x,y),\quad  \qquad &&  V(x,-y)=V(x,y)
\nonumber\\
\qquad\Phi^c(x, -y)=-\Phi^c(x,y) \qquad && \Sigma(x,-y)=-\Sigma(x,y),
\end{eqnarray}
where $\Phi$ can be any of the SM fields $Q, U, D, L, E$. 
As a result, the original 5D N=1 supersymmetry is broken
and  the  fixed points ($y=0, \pi R$)
of the orbifold have  a remaining
4D N=1 supersymmetry. Further, we consider
 the following localised interaction,

\begin{eqnarray}\label{interaction}
  \cL_4 = \int\!  dy \; \delta(y) 
  \left\{-\int \!\!d^2\! \theta 
 \,\, \Big[\lambda_t\, Q \,U \,H_u 
      + \lambda_b \, Q \, D\,  H_d + \cdots \Big]
  + {\rm h.c.} \right\}.\\[-11pt]
\nonumber
\end{eqnarray}
The 5D coupling 
$\lambda_{t}=f_{5, t}/M_*^n=(2\pi R)^n f_{4, t}$ where $f_{5,t}$ 
 ($f_{4, t}$) is  the dimensionless 5D  (4D), 
$M_*$ is the cutoff of the theory. 
In the following $Q$, $U$, $D$ superfields are {\it always}
bulk fields  and have mass  dimension 
$[Q]\!=\![U]\!=\![D]\!=\!3/2$, while the Higgs field $H_{u,d}$ can be a 
brane field $[H_{u,d}]=1$ if $n=1$  or
a bulk field\footnote{If $H_{u,d}$ are bulk fields they also
satisfy a condition similar to that for $\Phi$ in eq.(\ref{orbifold}).}
 $[H_{u,d}]=3/2$, when $n=3/2$ (when it also has a 
$H_{u,d}^c$ partner). The above spectrum and interactions define our 
minimal model. While this is not phenomenologically viable, 
it is sufficient to illustrate the idea of the paper. 
Moreover the spectrum and interaction considered are generic 
in many detailed  models which reproduce the SM or its 
supersymmetric versions. Such models, for which our findings are
relevant, can be found in  refs.\cite{Antoniadis} to 
\cite{DiClemente:2002qa}.

The purpose of this work is to investigate the correction to the mass of 
the brane (or  zero-mode of the bulk) scalar fields $\phi_{H_u}$, $\phi_{H_d}$ 
induced\footnote{One needs two Higgs fields of opposite
 hypercharge  to avoid  (quadratically divergent) 
FI corrections \cite{Ghilencea:2001bw}.} by towers of
Kaluza-Klein states of $Q, U, D$ superfields, 
via interaction~(\ref{interaction}). Gauge corrections will also be considered.
Since similar considerations apply to both $H_u$ and $H_d$, we shall 
compute the one-loop correction to the mass $m_{\phi_{H}}$ of 
$\phi_{H_u}$, hereafter denoted simply $\phi_H$.
 To compute the one-loop  corrected $m_{\phi_H}$
one must first evaluate  the 
spectrum of Kaluza-Klein modes, using eq.(\ref{orbifold}). However
this spectrum also  depends on the mechanism of the further breaking of
4D N=1 supersymmetry which is then transmitted to
the ``visible'' sector at $y=0$.  We shall distinguish 
two cases discussed separately:
{\bf I. }~Localised supersymmetry breaking  and 
{\bf II.}~Non-local supersymmetry breaking.

\subsection{Higher derivative operators
 from localised supersymmetry breaking.} \label{section2.1}

We  consider that supersymmetry is broken  at a distant (hidden) brane
located at $y=\pi R$ by
\begin{eqnarray}\label{ZZ}
\cL_4=\int dy \,\delta(y-\pi R)\, \bigg[
\int d^2\theta \,\,M_*^2 \, Z+h.c.\bigg],
\end{eqnarray}
where $Z$ is a (gauge singlet) brane field at $y=\pi R$.
The bulk  fields  $Q, U$ feel the   supersymmetry breaking via
the couplings
\begin{eqnarray}\label{break_susy}
\cL_4=\int dy\, \delta(y-\pi R) \left\{-\int d^4\theta \,\,\bigg[
\frac{c_Q}{M_*^3}\,Q^\dagger Q\,Z^\dagger Z+
\frac{c_U}{M_*^3}\,U^\dagger U\,Z^\dagger Z\bigg]\right\}.
\end{eqnarray}
Therefore, when $\langle Z\rangle\sim F_Z\theta^2$, 
the bulk fields such as $\phi_{Q,U}$ with non-zero coupling at the $y=\pi R$
brane have the spectrum modified.
One can show  that (for details see for example \cite{Arkani-Hamed:2001mi})
\begin{eqnarray}\label{eq7}
\tan [\pi R m_{\phi_{M},n}]=\frac{c_{M}}{2} \frac{F_Z^2}{M_*^4}
\frac{M_*}{m_{\phi_{M},n}}, \qquad \Rightarrow \qquad
m_{\phi_{M},n}= \Big(n+\frac{1}{2}\Big) \frac{u}{R},\qquad M=Q, U.
\end{eqnarray}
With the choice $c_M\sim \cO(1)$, $F_Z\sim M_*^2$, 
and to leading order in $1/(R M_*)$, one can set $u=1$. 
This is usually referred to as  ``strong'' supersymmetry breaking, 
otherwise $u$ is a series in $1/(R M_*)$.

Unlike the fields $\phi_{Q,U}$, their
fermionic partners $\psi_{Q, U}$ do not couple to the vev of $Z$, 
and their mass is not changed by (\ref{ZZ}).
This results in the  breaking at $y=\pi R$ of the remaining 
4D N=1 supersymmetry of zero modes. Similarly, the fields $\psi^c_{Q,U}$,
$\phi^c_{Q,U}$ do not couple to $Z$ (see eq.(\ref{orbifold})), 
thus their mass spectrum
is not affected. Finally, the Higgs field $H_u$ (also $H_d$)
is considered to be a {\it brane} field,  localised at $y=0$, so  it does
not have tree level interactions with the field $Z$.
However,  it feels the supersymmetry breaking at $y=\pi R$
through loops of the bulk  fields $Q, U$, (or $Q, D$ for $H_d$) 
via eq.(\ref{interaction}).
Therefore the
 spectrum of the bulk fields is, using (\ref{orbifold}), (\ref{eq7})
\begin{eqnarray}
m_{\psi_M,k} & = & \frac{k}{R}, \qquad\qquad  \quad k\geq 0,
 \qquad\qquad\quad  m_{\psi_M^c,k}=\frac{k}{R}, \quad k\geq 1
 \label{eqm1}
\nonumber\\
m_{\phi_M,k}  & = & u\bigg(\frac{k + r^{\phi}}{R}\bigg),\quad k\geq 0,
\qquad\qquad\quad m_{\phi_M^c,k}=\frac{k}{R},\quad  k\geq 1,\qquad
M\equiv Q, U.
\end{eqnarray}
where $r^\phi=1/2$ and $u=1$ for strong supersymmetry breaking. 
The case with $u\not\! =\! 1$ is discussed in Section \ref{section3}.
Finally,  the wavefunction normalisation coefficients (at $y\!=\!0$)
are $\eta_k^{\phi_M}\!=\!1$, $\eta_k^{\psi_M}\!=\!
\eta_k^{F_M}\!=\!1/{\sqrt 2}^{\delta_{k,0}}$ ($k\geq 0$). 
These results will be used in Section~\ref{section2.1.1} to compute 
$m_{\phi_H}$ at one-loop.

Finally, 
let us also consider the case of supersymmetry breaking at $y=\pi R$
with a   brane-localised  gaugino mass on $S^1/Z_2$,
\begin{eqnarray}\label{bk}
\cL_4=\int dy \,\delta(y-\pi R) \int d^2\theta \frac{1}{(4 g_5)^2}
\frac{c_W}{M_*^2}\, Z\, {\rm Tr}\, \Big[ W^\alpha W_\alpha \Big]+ h.c.
\end{eqnarray}
Here $W_\alpha$ contains gaugino as its lowest component.
Note that only the even parity gaugino $\lambda_1$ (see
eq.(\ref{orbifold})) couples to Z
and feels the supersymmetry breaking at $y=\pi R$. Further,
$\lambda_1$ and $\lambda_2$ are coupled through their kinetic term.
Eq.(\ref{bk}) is the counterpart of eq.(\ref{break_susy}) for the gauge case.
Using the equation of motion one finds  \cite{Arkani-Hamed:2001mi}
\begin{eqnarray}\label{eqw1}
\tan[\pi R m_\lambda]=\frac{c_W}{4 M_*^2} \, F_Z,
\end{eqnarray}
Then, the spectrum of gaugino is
\begin{eqnarray}
m_{\lambda_1}=m_{\lambda_2}=\frac{k+\rho}{R}, \ \ \  k\in {\bf Z}, 
\end{eqnarray}
with 
\begin{eqnarray}\label{eqw2}
\rho=\frac{1}{\pi}\,{\rm arctan}\bigg(\frac{c_WF_Z}{4 M^2_*}\bigg). 
\end{eqnarray}
This result is used in Section \ref{section2.1.2} for the 
one-loop gauge correction to the scalar mass $m_{\phi_H}$.

\subsubsection{One-loop mass correction due to Yukawa
  interaction}\label{section2.1.1}

\begin{figure}[t] 
\begin{center}
\begin{tabular}{cc|cr}
\parbox{6cm}{
\scalebox{0.83}{\mbox{
\psfig{figure=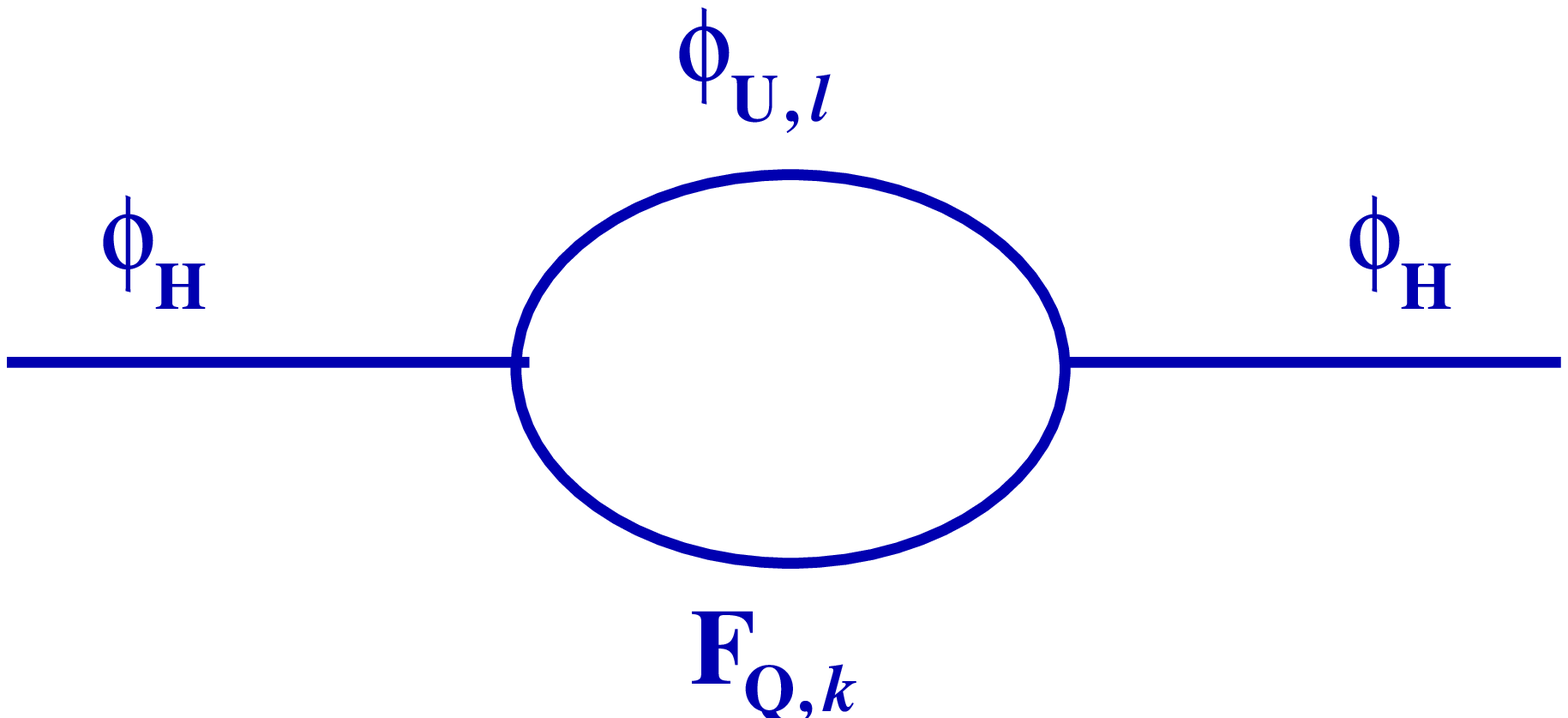,height=1.2in,width=2.5in,angle=0}}}}
\hspace{1.3cm}
\parbox{6cm}{\scalebox{0.83}{\mbox{
\psfig{figure=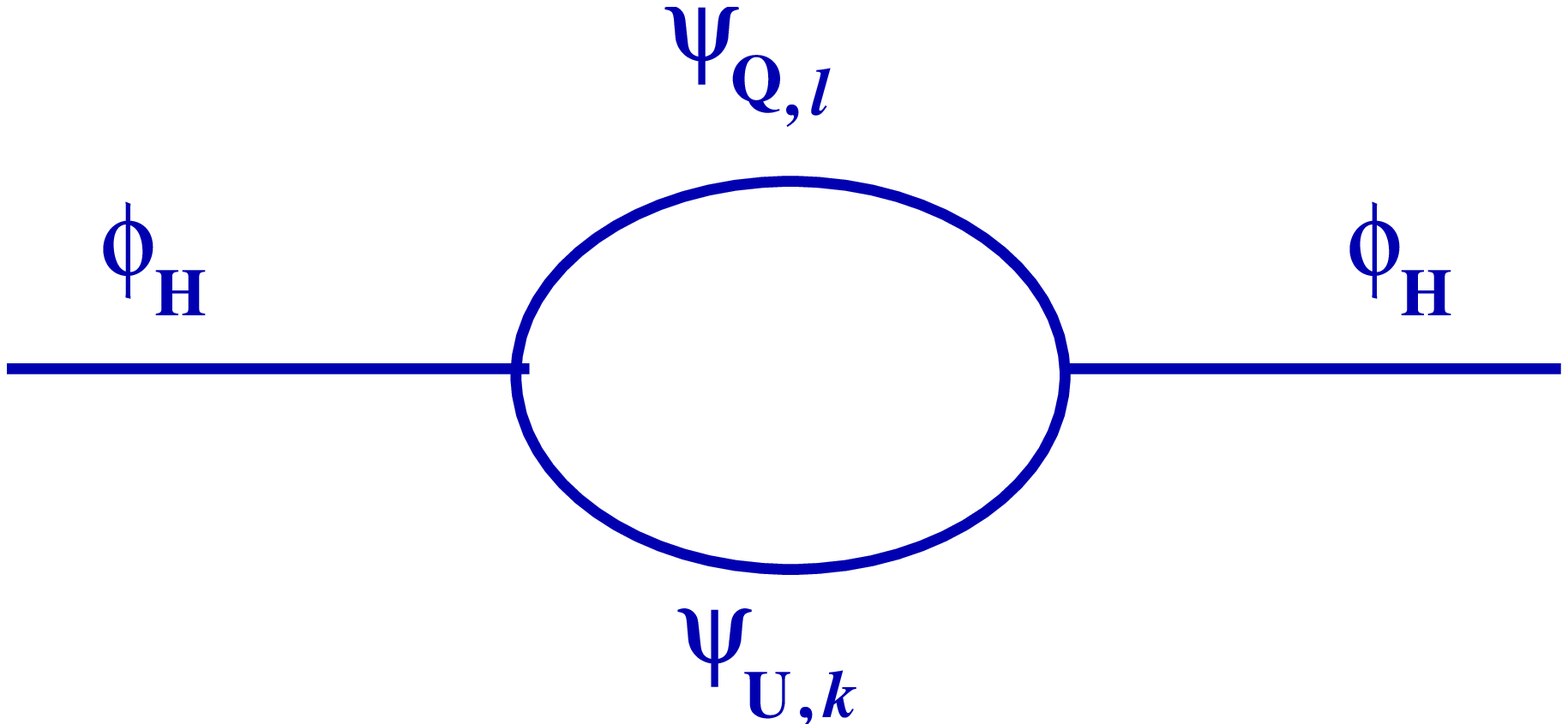,height=1.2in,width=2.5in,angle=0}}}}
\end{tabular}
\end{center}
\caption{\small{Diagrams contributing to the two-point Green function of the
 (brane or zero mode of the bulk) 
scalar field $\phi_H$. For the left diagram one should also  add the
 similar contribution with $Q\leftrightarrow U$.}}
\label{fig111}
\end{figure}

In the on-shell formulation, the interaction in eq.(\ref{interaction}) 
becomes
in component fields \cite{Arkani-Hamed:2001mi}

\begin{eqnarray}\label{action4d}
  \cL_4 &=&
    \sum_{k=1}^{\infty} \sum_{l=0}^{\infty} 
     \,(2 \,f_{4, t})\, \Bigl[ \, m_{\phi_Q^c,k} \,
    \eta^{F_Q}_k \eta^{\phi_U}_l\,
    \phi^{c\,\dagger}_{Q,k} \,\phi_{U,l}\, \phi_H
 + {\rm h.c.} + (Q \leftrightarrow U) \Bigr]
\nonumber\\[3pt]
&-& \sum_{k=0}^{\infty} \sum_{l=0}^{\infty} \sum_{m=0}^{\infty} 
     (2 \,f_{4, t})^2\, \Bigl[
    \eta^{\phi_Q}_k \eta^{\phi_Q}_l (\eta^{F_U}_m)^2
    \phi^{\dagger}_{Q,k} \phi_{Q,l} \phi^{\dagger}_H \phi_H
+ (Q \leftrightarrow U)
\Bigr]
\nonumber\\[3pt]
&-& \sum_{k=0}^{\infty} \sum_{l=0}^{\infty} 
  \,(2 \,f_{4, t})  \Bigl[ \eta^{\psi_Q}_k \eta^{\psi_U}_l
    \psi_{Q,k} \psi_{U,l} \phi_H 
  + {\rm h.c.} \Bigr]. \\
\nonumber
\end{eqnarray}
Note that the sum over $k$ in the first line is from $k\geq 1$ since
the field $\phi^c$ is odd under orbifolding, eq.(\ref{orbifold}).
If we work on-shell (off-shell) there are three (two) types of 
diagrams contributing to the mass of brane Higgs field $\phi_H$, see 
Fig.~\ref{fig111}. The one-loop contributions in Euclidean space 
are\footnote{This form of the  radiative corrections 
can be shown to be equivalent to that derived using 
5D Green functions in mixed positions-momentum space 
(computed in ref.\cite{Arkani-Hamed:2001mi}, eqs.(57), (60), (64)),
and evaluated at $y=0$. This also suggests the (brane) localisation
of the corresponding  counterterm, see later.}

\begin{eqnarray}\label{mass}
-i\, m_{\phi_H}^2 (q^2)\bigg\vert_B\!\!\!\! \!&=& \!\!\!\!
i \,  (2 f_{4, t})^2 N_c\!\!\! \sum_{k\geq 0,\, l\geq 0}
    \Big[\eta^{F_Q}_k \eta^{\phi_U}_l\Big]^2 
   \!\! \int \frac{d^d p}{(2\pi)^d} 
\frac{(-1)(p+q)^2\,\,\mu^{4-d}}{((p+q)^2+m_{\phi_Q^c,k}^2)(p^2+m_{\phi_U,l}^2)}
    \!+\! (Q\! \leftrightarrow\! U)
\nonumber\\[12pt]
-i \, m_{\phi_H}^2 (q^2)\bigg\vert_F\!\!\!\!\!
 &= &\!\!\!\!  i  \,(2 f_{4, t})^2 N_c \!\sum_{k\geq 0,\, l\geq 0}
    \Big[\eta^{\psi_Q}_k \eta^{\psi_U}_l\Big]^2
    \!\int \frac{d^d p}{(2\pi)^d} 
    \frac{2 \,p.(p+q)\,\,\mu^{4-d}}{((p+q)^2
+m_{\psi_Q,k}^2)(p^2+m_{\psi_U,l}^2)}\\
\nonumber
\end{eqnarray}

\noindent
Here $d\!=\!4\!-\!\epsilon$  ($\epsilon\!\ra\! 0$), $N_c$ is the
number of colours, $\mu$ is the scale
introduced by the DR scheme, $q^2$ is the external momentum and
the two double sums extend to infinity. The index $B, (F)$ 
stands for bosonic (fermionic) contributions. One 
then uses the spectrum (\ref{eqm1}) 
and coefficients $\eta$ given after this equation, 
perform the  integrals over $p$ in DR, then the double sums,  to find

\begin{eqnarray}\label{mass1extra}
-\, m_{\phi_H}^2 (q^2)\bigg\vert_B\!\!\!\! \!&=& \!\!\!\!
  (2 f_{4, t})^2\, N_c\!\!\! \sum_{k\geq 0,\, l\geq 0}
\frac{1}{2^{\delta_{k, 0}}}
 \!\! \int \frac{d^d p}{(2\pi)^d} 
\frac{(-2) (p+q)^2\,\,\mu^{4-d}}{((p+q)^2+k^2/R^2)(p^2+(l+1/2)^2/R^2)}
\nonumber\\[10pt]
&=&
-\frac{(2 f_{4,t})^2 \,\kappa_\epsilon}{2\,(4 \pi R)^2} N_c
\int_0^1 dx\,\bigg\{
\frac{2-\epsilon/2}{\pi}\,\cJ_2[1/2,0, c]+ q^2 R^2 (1-x)^2
  \,\cJ_1[1/2,0, c]\bigg\}
\nonumber\\[10pt]
-\, m_{\phi_H}^2 (q^2)\bigg\vert_F\!\!\!\!\!
 &= &\!\!\!\! (2 f_{4, t})^2\, N_c \!\!\! 
\sum_{k\geq 0,\, l\geq 0}
  \frac{1}{2^{\delta_{k,0}}} \frac{1}{2^{\delta_{l,0}}}
 \!\int \frac{d^d p}{(2\pi)^d} 
    \frac{2 \,p.(p+q)\,\,\mu^{4-d}}{((p+q)^2+k^2/R^2)(p^2+l^2/R^2)}
\nonumber\\[10pt]
&=&
\frac{(2 f_{4,t})^2 \,\kappa_\epsilon}{2\, (4 \pi R)^2} \,N_c
\int_0^1 dx\,
\bigg\{\frac{2-\epsilon/2}{\pi}\,\cJ_2[0,0, c]+ q^2 R^2 x (x-1)
  \,\cJ_1[0,0, c]\bigg\},\\
\nonumber
\end{eqnarray}
where $\kappa_\epsilon\equiv (2 \pi \mu R)^\epsilon$. 
The functions $\cJ_{1,2}$ have the following definition and leading
behaviour in~$\epsilon$

\begin{eqnarray}\label{j12functions}
\cJ_j[ c_1,c_2,c]\!\! &\equiv &\!\!\!\!\!\!\sum_{k_1,k_2\in \bZ}\int_0^\infty
\!\!\!\frac{dt}{t^{j-\epsilon/2}}\,\, e^{-\pi\,
    t\,(c+a_1(k_1+c_1)^2+a_2(k_2+c_2)^2)}=
 \frac{\big(\!-\!\pi  c\,\big)^j}{j \sqrt{a_1 a_2}}\,
\bigg[\frac{2}{\epsilon}\bigg]\!+\cO(\epsilon^0),\,\,\,\,\, j\!=\!1,2.
\nonumber\\
\nonumber\\
&& a_1=(1-x), \qquad\, a_2= x, \qquad\, c=x (1-x) \,q^2 R^2.
\\[-9pt]
\nonumber
\end{eqnarray}
For a complete expression of the functions $\cJ_{1,2}$ see the
Appendix,
eqs.(\ref{j1general}) to (\ref{a3}) and (\ref{pp1}) to (\ref{pp2}).
It is important to notice that the leading (divergent) behaviour of
$\cJ_{1,2}$ depends on $c$ and $a_1, a_2$ but is independent of
$c_1$, $c_2$. The dependence on $c$ is very important, since it is
only for $c\sim q^2 R^2 \not=0$ i.e. non-zero external momentum
that one is able to ``see'' the poles of
$\cJ_{1,2}$. After adding the bosonic and fermionic contributions 
in (\ref{mass1extra}), one finds

\begin{eqnarray}\label{mass3}
-\, m_{\phi_H}^2 (q^2)
&=& \frac{(2 f_{4,t})^2}{2\, (4\pi R)^2} \, N_c\, \bigg\{\!
\int_0^1  dx \, 
(2/\pi)\,\Big[ \cJ_2[0,0,c]-\cJ_2[1/2,0,c]\Big]
\nonumber\\[12pt]
&+&\,
\kappa_\epsilon (q^2 R^2)
\int_0^1  dx \Big[\,x \,(x-1)\,
\cJ_1[0,0,c]-(1-x)^2\,\cJ_1[1/2,0,c]\Big]\!\bigg\}
\end{eqnarray}
with $c$ as in (\ref{j12functions}).
Note that if $q^2=0$ the second line above is
 absent, so $m^2_{\phi_H}(q^2=0)$ is
given by the first line alone. Further, in the difference
 $\cJ_2[0,0,c]-\cJ_2[1/2,0,c]$ the divergent part $q^4 R^2/\epsilon$
in each $\cJ_2$ cancels away to give a one-loop finite $m^2_{\phi_H}(0)\sim 1/R^2$.
This cancellation is ensured
by the equal number of bosonic and fermionic degrees of freedom, 
enforced by the initial supersymmetry.
Using the leading behaviour of the functions
$\cJ_{1,2}$ we find
\begin{eqnarray}\label{mass_final}
m_{\phi_H}^2(q^2)=m_{\phi_H}^2(0)- 
\frac{(2 f_{4,t})^2}{2^8} 
N_c\, (q^4 R^2)\, \bigg[\frac{1}{\epsilon}+\ln(2\pi R \mu)\bigg]
 +\frac{1}{R^2} \,\,\cO(q^2 R^2)
\end{eqnarray}
where the $\mu$-dependent  term in the square bracket shows
the regularisation scheme dependence induced by 
$\kappa_\epsilon$. The finite part
$\cO(q^2 R^2)$ can be evaluated from the second line in (\ref{mass3}) 
using eqs.(\ref{a2}), (\ref{a3}) in
the Appendix to give the full running of the mass wrt momentum scale
$q^2$.

Eq.(\ref{mass_final}) shows the presence
 in the sum of bosonic and fermionic contributions,
of a pole multiplied by
 quartic dependence on (external momentum)
 $q$, originating from the two $\cJ_1$'s. The
result is that the one loop scalar mass is not finite and has a UV
divergence similar to that cancelled by (initial) supersymmetry in the
 $J_2$ dependent part.  
Therefore one
must add in the action a  higher derivative counterterm\footnote{This
result is similar to that in \cite{Ghilencea:2004sq} which had instead
5D N=1 supersymmetry broken to N=0 on $S_1/(Z_2\!\times\!Z_2')$.}
to $m^2_{\phi_{H}}$ (recalling that $\phi_H$ is a brane field) 
\begin{eqnarray}\label{hdo0}
\int \! d^4 x \,d^2 \theta \, d^2\overline \theta 
\lambda^2_t \, H_u^\dagger \Box  H_u  
\sim \! f_{4,t}^2\!
\int \!\! d^4 x \,
 R^2 \, \phi_{H}^\dagger  \Box^2  \phi_{H}
+...
\end{eqnarray}
The presence of the UV divergence and of corresponding
higher derivative counterterm shows that,
although the initial theory was supersymmetric, its non-renormalisable
character is nevertheless manifest through a  counterterm generated
by the large number (multiplicity) of Kaluza-Klein modes
which contribute to $m_{\phi_H}$.

\begin{table}[tbp] 
\begin{center}
\begin{tabular}{|l|l|l|l|c|} 
\hline
$[Q]$  & $[U]$   & $[H_u]$ &  $[\lambda]$  & $\#$ (no. of loops) \\[4pt]
\hline\hline 
3/2 (bulk) &   3/2 (bulk) & 1  (brane)  &   -1          &  $n=1$          \\  
3/2  &   1   & 3/2 &   -1          &  $n=1$           \\
3/2  &   3/2 & 3/2 &   -3/2        &  $n=1$           \\
3/2  &   1   & 1   &   -1/2        &  $n=2$           \\  
1    &   1   & 3/2 &   -1/2        &  $n=3$           \\
\hline
\end{tabular} 
\end{center} 
\caption{\small This is an {\it estimate} of the number of loops $n$
when higher derivative operators may be
 generated. The localised superpotential 
$\int dy\, \delta(y)\, d^2\theta\,\lambda \,Q U H_u$ 
with $Q, U, H_u$ as  brane/bulk fields
 generates higher derivative counterterms to the scalar zero mode
 of $H_u$ (if a bulk field) or scalar component of $H_u$ (if a brane field).
An example is $\int d^4\theta \,(\lambda^2)^n H_u \Box H_u$. 
The table is a dimensional estimate of  the number  of loops 
when this counterterm arises located at $y=0$,  
in function of the nature (bulk/brane) 
of the fields.}\label{table:1} 
\end{table} 
 
In Table \ref{table:1} we provided a general dimensional analysis for
when higher derivative counterterms can arise from a
localised superpotential. The table shows the number of loops in 
perturbation theory at which such counterterms can be generated,
by assuming that any of the fields in the superpotential can be 
bulk or brane fields. 
This information can be used in realistic orbifold models, to
avoid such operators at low orders in perturbation theory.

\subsubsection{One-loop gauge correction to a brane scalar mass}
\label{section2.1.2}

Let us now consider the one-loop
gauge contribution to the  mass $m_{\phi_H}$
of the brane scalar $\phi_H$ located at $y=0$, 
assumed to be charged under a $U(1)$ group. This is induced by the
action (\ref{bk}) and in the following we use eqs.(\ref{eqw1}) to 
(\ref{eqw2}).
In the dimensional regularisation with $d=4-\epsilon$, 
bosonic (gauge) and fermionic (gauginos)
 contributions to the scalar self-energy 
at nonzero external momentum $q^2$, are respectively
\bea
-i\, m^2_{\phi_H}(q^2)\Big\vert_B=(-i)\, 4\, g^2_4 \,
 \mu^{4-d}\sum_{n\in {\bf Z}}\int 
\frac{d^d p}{(2\pi)^d}\frac{p.(q+p)}{(p^2+n^2/R^2)(q+p)^2} 
\eea
and
\bea
-i\, m^2_{\phi_H}(q^2)\Big\vert_F=i \,4\, g^2_4\, \mu^{4-d}\sum_{n\in {\bf Z}}\int
\frac{d^d p}{(2\pi)^d}\frac{p.(q+p)}{(p^2+(n+\rho)^2/R^2)(q+p)^2}. 
\eea

\noindent
Then, we find the one-loop corrections as

\bea
m^2_{\phi_H}(q^2)\Big\vert_B
=\frac{g^2_4(\mu\pi R)^\epsilon}{4\pi^3R^2}\int^1_0 dx
\bigg[\Big(2-\frac{\epsilon}{2}\Big)\, {\cal G}_2[0,c]
-\pi x(1-x)\, q^2 R^2\, {\cal G}_1[0,c]\bigg]
\eea
and
\bea
m^2_{\phi_H}(q^2)\Big\vert_F
=-\frac{g^2_4(\mu \pi R)^\epsilon}{4\pi^3R^2}\int^1_0 dx
\bigg[\Big(2-\frac{\epsilon}{2}\Big)\, {\cal G}_2[\rho,c] 
-\pi x(1-x)\, q^2 R^2\, {\cal G}_1[\rho,c]\bigg]
\eea

\noindent
with $c\equiv x(1-x)q^2R^2$ and we introduced

\bea\label{ss1}
{\cal G}_j[\rho,c]\equiv \sum_{n\in {\bf Z}}
\int^\infty_0 \frac{dt}{t^{j-\epsilon/2}}
e^{-\pi t[c+\beta(n+\rho)^2]}, \ \ j=1,2; \ \ \beta=x.
\eea
After some calculations one obtains \cite{Ghilencea:2003xy}
\begin{eqnarray}\label{gc1}
\cG_1[\rho,c]&=&  -\ln\bigg\vert2 \sin \pi (\rho+i
\sqrt{c/\beta}\,)\bigg\vert^2,
\nonumber\\[12pt]
 \cG_2[\rho,c] 
&=& \frac{4\pi^2 \,c^{3/2}}{3 \,\beta^{1/2}}
+\bigg\{ (\beta\,c)^{{1}/{2}}\,
\Li_2\Big(e^{2 i \pi (\rho+i
  \sqrt{c/\beta}\,)}\Big)+\frac{\beta}{2\pi}\,\Li_3\Big(
e^{2 i \pi (\rho+i  \sqrt{c/\beta}\,)}\Big)+c.c.\bigg\}
\end{eqnarray}

\noindent
where $c,\,\beta>0$ and 
$\Li_\sigma(x)=\sum_{k\geq 1} x^k/k^\sigma$. 
Although these
 expressions are finite (no poles in $\epsilon$), 
whenever one removes a mode 
from the series, poles arise and this justifies 
keeping the $\epsilon$ dependence explicit in their definition eq.(\ref{ss1}).

Therefore, the resulting one-loop correction for the brane scalar
is given by
\bea\label{eq26}
m^2_{\phi_H}(q^2)&=&m^2_{\phi_H}(q^2)\Big\vert_B+m^2_{\phi_H}(q^2)\Big\vert_F 
=\frac{g^2_4}{2\pi^3 R^2}\int^1_0 dx\bigg[{\cal G}_2[0,c]
-{\cal G}_2[\rho,c]\bigg] \nonumber \\[8pt]
&&-\frac{g^2_4}{4\pi^2 R^2}\, (q^2\, R^2)\int^1_0 dx\, x(1-x)
\bigg[{\cal G}_1[0,c]-{\cal G}_1[\rho,c]\bigg].
\eea

\noindent
This result can be further simplified, but for our purpose it is
enough to notice that it is finite (has no poles in $\epsilon$). Therefore
we conclude that
no higher derivative counterterms are generated  at the one-loop level.
We note that the momentum-independent mass correction is given by
\begin{eqnarray}\label{eq27}
m^2_{\phi_H}(0)=
\frac{g^2_4}{4\pi^4 R^2} \bigg[\zeta[3]- \frac{1}{2}\Big(
\Li_3(e^{2 i \pi \rho})+c.c.\Big)\bigg]
\end{eqnarray}

\noindent
This result agrees with the one obtained by changing 
the infinite Kaluza-Klein sum into a contour integral \cite{choi-lee}.

Alternatively, one can investigate the gauge correction
 by considering supergraphs.
Given the presence of subtle differences between 
component and supergraph formalisms  \footnote{related to  
gauge fixing in the WZ gauge in component formalism as compared to 
superfield gauge fixing in~5D.}   we now use the supergraph formalism to 
show again that no higher derivative counterterms are present. This will
confirm our conclusion obtained  in the component  formalism.

For this purpose we compute
 the gauge correction to the propagator of a (massless) brane
chiral multiplet $H$ in the absence of supersymmetry breaking.
To do so we need to consider only one supergraph with brane-chiral 
and bulk-vector multiplets ``running'' in the loop \cite{Grisaru}.
We assume that, as in the 4D case, the soft breaking does not renormalise 
the propagator of a massless brane chiral multiplet.
With an appropriate gauge fixing term \cite{Flacke},
i.e. the 5d version of the super Feynman gauge,
the action for the vector superfield is
\be
{\cal L}_5=\int d^5x d^2\theta d^2{\bar\theta}\,V[-\Box-\partial^2_5] V.
\ee
Thus, the propagator for the bulk vector multiplet on $S^1/Z_2$ satisfies
\be
(-q^2+\partial^2_5)\Delta_5(y-y',\theta-\theta')
=-\frac{1}{2} \sum_{n\in {\bf Z}}\delta(y-y'-2\pi n R)
\,\,\delta^4(\theta-\theta').
\ee
Therefore, for $0\leq y-y' \leq \pi R$,
the mixed position-momentum  propagator is given by
\begin{eqnarray}
\Delta_5=-\frac{1}{2}\,G_5(y,y')\delta^4(\theta-\theta')
\qquad {\rm {with}} \qquad
\qquad
G_5(y,y')=\frac{\cosh[ q\,(y-y'-\pi R)]}{2\, q\sinh(\pi R\,q)}.
\end{eqnarray}

\noindent
In particular, the bosonic part of the propagator at the origin
with $y=y'=0$ is given by
\bea
G_5(0,0)&=&\frac{1}{2 q \tanh(\pi R\,q)}
=\frac{1}{2\pi R}\sum_{n\in {\bf Z}}\frac{1}{q^2+(n/R)^2}.
\eea

\noindent
We obtain the one-loop gauge correction to the propagator
of the brane chiral multiplet located at $y=0$ as
\be\label{eq32}
-\frac{g^2_5}{2\pi R}\int \frac{d^4 q}{(2\pi)^4}A(q)\bigg[\int d^4
\theta\, {\overline H}(-q,\theta)\, H(q,\theta)\bigg]
\ee
where
\bea
A(q)&=&\mu^{4-d}\sum_{n\in {\bf Z}}
\int\frac{d^dk}{(2\pi)^d}\frac{1}{(q+k)^2(k^2+n^2/R^2)} \nonumber \\[9pt]
&=&\frac{1}{(4\pi)^{d/2}(\pi\mu^2 R^2)^{d/2-2}}\int^1_0dx
\sum_{n\in {\bf Z}}\int^\infty_0\frac{dt}{t^{d/2-1}}e^{-\pi t(c+xn^2)}
\eea

\noindent
where $c=x(1-x)\,q^2 R^2$. 
Finally, with $d=4-\epsilon$ and using eqs.(\ref{ss1}), (\ref{gc1}),
one finds that the  one-loop gauge correction
to the kinetic term of the brane scalar $\phi_H$ (component of $H$) 
in (\ref{eq32}) has a momentum dependence
of type
\begin{eqnarray}
q^2\,A(q)=-\frac{1}{(4\pi)^2}\frac{1}{(\pi R)^2}\bigg[\frac{4\,y^3}{3} 
- \zeta[3]+ 2\, y\,\,\Li_2\big(e^{-2 y}\big)
+\Li_3\big(e^{-2 y}\big)\bigg]+
\cO(\epsilon),\quad y=\pi\,R\,\sqrt{q^2}.
\end{eqnarray}

\noindent
Since the above result  has no poles in $\epsilon$,
we find, using this time  the supergraph computation, 
that the wave function of the brane chiral multiplet
is not renormalised. Therefore no higher 
derivative  counterterms arise at one-loop order, in agreement with the
previous computation using the component field formalism.

Finally, let us discuss the possible higher dimensional 
(derivative) operators  which can be induced by the gauge 
corrections at higher loops, along the lines 
discussed in Table~\ref{table:1}.
The corresponding operator for a brane chiral multiplet is
$$\int d^4\theta (g^2)^n \, {\bar H}\Box H$$
 with $n$ being the number of loops.
Then, given the (mass) dimension of the higher dimensional gauge coupling,
i.e. $[g_5^2]=-1$ in 5D  higher dimensional operators can in principle
 be generated for $n=2$ (two-loop). However,  for a 6D case, $[g_6^2]=-2$, 
higher derivative  operators may in principle be 
generated for $n=1$ (one-loop) \footnote{A component
field computation on $T^2/{\bf Z}_2$ shows that one-loop gauge correction
to the self-energy of a brane scalar is finite \cite{Lee}.
The agreement of this result with that using   a  supergraph 
approach is studied elsewhere~\cite{Hyun}.}.

\subsection{Higher derivative operators
  from non-local  Supersymmetry breaking.}\label{section2.2}

So far we have considered a brane-localised supersymmetry breaking,
eqs.(\ref{ZZ}), (\ref{break_susy}), (\ref{bk}).
In the following we consider a non-local supersymmetry breaking 
mechanism, with: (1) discrete and (2) 
continuous twisted boundary  conditions, which we examine separately.

\subsubsection{Discrete Scherk-Schwarz  twists.}\label{2.2.1}

First, let us impose that
the 5D fields acquire  under a $2\pi R$ shift, a phase
which is the $R$-parity charge of these fields. 
The action of the $R$-parity operator is
\begin{eqnarray}\label{rparity}
Z_{2,R} Q(x,y,\theta)    =-Q(x,y,-\theta),\qquad &&
Z_{2,R} Q^c(x,y,\theta)  =-Q^c(x,y,-\theta),\nonumber\\
Z_{2,R} H(x,y,\theta)    =H(x,y,-\theta),\quad\qquad\! && \!
Z_{2,R} H^c(x,y,\theta)  =H^c(x,y,-\theta),\nonumber\\
Z_{2,R} V(x,y,\theta)    =V(x,y,-\theta),\qquad\quad &&\,\,\,
Z_{2,R}\Sigma(x,y,\theta)=\Sigma(x,y,-\theta)
\end{eqnarray}
One also has a condition for $U, U^c$ superfields similar to that for
$Q, Q^c$. The condition for $H, H^c$ stands for both
 $H_{u,d}, H^c_{u,d}$ and applies  only if these
fields  are bulk fields. Eqs.(\ref{rparity}), (\ref{orbifold})
give the spectrum relevant for our purpose 
\begin{eqnarray}
m_{\psi_M,k} & = & \frac{k}{R}, \qquad\quad\,\, k\geq 0,
 \qquad\qquad  m_{\psi^c_M,k}=\frac{k}{R}, \qquad\qquad k\geq 1 
\nonumber\\
m_{\phi_M,k}  & = & \frac{k + 1/2}{R},\quad k\geq 0, 
\qquad\qquad\,\, m_{\phi^c_M,k}=\frac{k+1/2}{R},\quad  
k\geq 0;\quad M=Q, U.
\label{eqq2}
\end{eqnarray}
and $\eta_k^{F_M}=\eta_k^{\phi_M}=1$ and 
$\eta_k^{\psi_M}=1/{\sqrt 2}^{\delta_{k,0}}$.
Also if the Higgs field is a bulk field, 
$m_{\phi_{H,k}}=k/R$ ($k\geq 0$) and $m_{\phi_{H^c, k}}=k/R$ ($k\geq
1$). Finally $m_{A_\mu,k}=k/R$, $m_{\lambda_{1,2,k}}=(k+1/2)/R$ 
($k\geq 0$).

Note  that the spectrum on the orbifold $S_1/Z_2$  with the R-parity 
(\ref{rparity})
has similarities with that  in  the case of  $S_1/(Z_2\times Z_2')$ orbifold  
(see for example \cite{Barbieri:2000vh}) with
the $Z_2'$ identified with a $Z_{2,R}$ Scherk-Schwarz breaking of
supersymmetry. As a result the one-loop corrected  $m_{\phi_H}(q^2)$
is expected to be similar to that in $S_1/(Z_2\times Z_2')$
studied in\footnote{For the completeness of our analysis we 
include this case  here in detail.} \cite{Ghilencea:2004sq}. 
This similarity is only present when considering  interactions
from one  fixed point, and breaks down when 
overlapping interactions from different fixed points are included,
as can happen at higher loops.

The action is similar to that in eq.(\ref{action4d})
with the remark that the first sum over $k$ starts from $k=0$.
With this information 
we can compute the one-loop corrections to the mass of the scalar
component  $\phi_H$ of $H_u$, or of its zero mode if $H_u$ is a bulk field.
One has in this case a result similar to eq.(\ref{mass}), but note that
the wavefunction coefficients  $\eta^{F_M}$ and the
 spectrum ($m_{\phi_Q^c}$) have  changed.
In Euclidean space

\begin{eqnarray}\label{mass2}
-i\, m_{\phi_H}^2 (q^2)\bigg\vert_B\!\!\!\! \!&=& \!\!\!\!
- i \,  f_t^2 N_c\!\!\! \sum_{k\geq 0,\, l\geq 0}
    \Big[\eta^{F_Q}_k \eta^{\phi_U}_l\Big]^2 
   \!\! \int \frac{d^d p}{(2\pi)^4} 
    \frac{2\,(p+q)^2 \,\,
\mu^{4-d}}{((p+q)^2+m_{\phi_Q^c,k}^2)(p^2+m_{\phi_U,l}^2)},
\nonumber\\[12pt]
-i \, m_{\phi_H}^2 (q^2)\bigg\vert_F\!\!\!\!\!
 &= &\!\!\!\!  i  \,f_t^2 N_c \!\sum_{k\geq 0,\, l\geq 0}
    \Big[\eta^{\psi_Q}_k \eta^{\psi_U}_l\Big]^2
    \!\int \frac{d^d p}{(2\pi)^4} 
    \frac{2\,p.(p+q)\,\,\mu^{4-d}}{
((p+q)^2+m_{\psi_Q,k}^2)(p^2+m_{\psi_U,l}^2)},
\\[-3pt]
\nonumber
\end{eqnarray}
with the notation 
$f_t\equiv 2^{n} f_{4, t}$, where $\,n=1$ ($n=3/2$) is $H_u$ is a brane (bulk)
field\footnote{If $H_u$ is a bulk field, the above correction
refers to its zero mode scalar.}. In  the first equation we used that 
the masses of  $\phi_Q$, $\phi_Q^c$, $\phi_U$,
$\phi_U^c$ are all equal, and this explains the presence of a 
factor 2 in the numerator of the integrand. 

The calculation
of eq.(\ref{mass2}) with replacements (\ref{eqq2}),
proceeds  as in Section \ref{section2.1.1}
(see also \cite{Ghilencea:2004sq}). After adding the bosonic and
fermionic contributions, one obtains 

\begin{eqnarray}\label{div}
-m_{\phi_{H}}^2 (q^2) & = & 
\frac{f_t^2}{16 \pi^3 R^2}N_c\int_0^1 dx\, 
\Big[ \cJ_2[0,0,c]-  \cJ_2[{1}/{2},{1}/{2},c] \Big]
\nonumber\\
\nonumber\\
&+& \frac{f_t^2}{32\, \pi^2 } \,\,\kappa_\epsilon\,
N_cq^2 \int_0^1  \! dx  \Big[
x(x - 1) \cJ_1[0,0,c]
-(1-\!x)^2 \cJ_1[{1}/{2},{1}/{2},c] \Big]
\\[-5pt]
 \nonumber
\end{eqnarray}
The terms involving $\cJ_j[1/2,1/2,c]$, $j=1,2$ account for the
bosonic contribution, while $\cJ_j[0,0,c]$  account for the
fermionic  part. The functions $\cJ_{1,2}$ are  given in
eq.(\ref{j12functions}). Comparing (\ref{div}) to
(\ref{mass3}) one notices a similar structure, but 
there is a difference in the arguments of 
$\cJ_{1,2}$ in the two equations. Since the pole structure 
of $\cJ_{1,2}[c_1,c_2,c]$ does
not depend on  the arguments $c_1$, $c_2$, the discussion of the UV 
divergences is not changed from that of eq.(\ref{mass3}).
Eqs.(\ref{div}), (\ref{j12functions}) give again
\begin{eqnarray}\label{dr}
m_{\phi_{H}}^2(q^2)=m_{\phi_{H}}^2(0)- 
\frac{f_t^2}{2^{8}}\,N_c\,  q^4\,R^2\bigg[\frac{1}{\epsilon}+\ln(2\pi
  R\mu)\bigg]
+  \frac{1}{R^2}\, \cO(q^2 R^2)
\end{eqnarray}
where $\cO(q^2 R^2)$ terms are due to $\cJ_1$ functions
and can  be evaluated numerically using the full expression of $J_1$
given in the Appendix.
The above result looks similar to that in (\ref{mass_final}), but
 the exact expression of $m^2_{\phi_H}(q^2)$ is different due to 
$\cJ_1$'s of different arguments.
The counterterm has then the structure (when $H_u$ is a bulk field)
\begin{eqnarray}\label{hdo1}
\int \! d^4 x \,d y \!\int \! d^2 \theta \, d^2\overline \theta
\,\delta(y)\, \lambda_t^2 \, H_u^\dagger \Box  H_u  
&\sim& \! f_t^2\!
\int\! d^4 x \,
 R^2 \!\!\! \sum_{n,p\geq 0} \phi_{H,n}^\dagger \Box^2  \phi_{H,p}
\nonumber\\
\nonumber\\
&\sim& \! f_t^2\!
\int \!\! d^4 x \,
 R^2 \, \phi_{H,0}^\dagger  \Box^2  \phi_{H,0}
+\cdots
\end{eqnarray}
with $[\lambda_t]=-3/2$.
If $H_u$ is a brane field instead ($[H_u]=1$ and 
$[\lambda_t]=-1$), the counterterm reads
\begin{eqnarray}\label{hdo2}
\int \! d^4 x \,d^2 \theta \, d^2\overline \theta \,\,
\lambda_t^2 \, H_u^\dagger \Box  H_u  
\sim \! f_t^2\!
\int \!\! d^4 x \,
 R^2 \, \phi_{H}^\dagger  \Box^2  \phi_{H}
+...
\end{eqnarray}

\subsubsection{Continuous Scherk-Schwarz  twists.}\label{section2.2.2}

Instead of a discrete twist eq.(\ref{rparity}),  in this Section we
 impose continuous twists on bulk fields by using the $SU(2)_R$
global symmetry. The $SU(2)_R$ action under 
$y\rightarrow y+2\pi R$ is
\begin{eqnarray}
\left(\begin{array}{l}\lambda_1 \\ \lambda_2\end{array}\right)(x,y+2\pi R)
&=&e^{-2\pi i\omega\sigma_2}\left(\begin{array}{l}\lambda_1 \\ 
\lambda_2\end{array}\right)(x,y), \\   
A_N(x,y+2\pi R)&=&A_N(x,y),\,\,\,N=\mu, 5. \\[10pt]
\left(\begin{array}{l}\phi_M \\ 
\phi^{c\dagger}_M \end{array}\right)(x,y+2\pi R)
&=&e^{-2\pi i\omega\sigma_2}\left(\begin{array}{l}\phi_M \\ 
\phi^{c\dagger}_M \end{array}\right)(x,y), \\   
\left(\begin{array}{l}\psi_M \\ \psi^{c\dagger}_M 
\end{array}\right)(x,y+2\pi R)
&=&\left(\begin{array}{l}\psi_M \\ \psi^{c\dagger}_M
\end{array}\right)(x,y),\qquad M\equiv Q, U.
\end{eqnarray}
where $(\phi_M,\phi^c_M)$ and $(\psi_M,\psi^c_M)$ are bulk quark
multiplets, $M=Q, U$. 
If we also allow the 
 Higgs multiplet(s) to live in the bulk, its boundary condition is 
\begin{eqnarray}
\left(\begin{array}{l}\phi_H \\ \phi^{c\dagger}_H 
\end{array}\right)(x,y+2\pi R)
&=&e^{-2\pi i\omega\sigma_2}\left(\begin{array}{l}\phi_H \\
\phi^{c\dagger}_H \end{array}\right)(x,y), \\ 
\left(\begin{array}{l}\psi_H \\ \psi^{c\dagger}_H 
\end{array}\right)(x,y+2\pi R)
&=&-\left(\begin{array}{l}\psi_H \\  
\psi^{c\dagger}_H \end{array}\right)(x,y)
\end{eqnarray} 
where we note that the higgsinos acquire only a phase of $R$-parity 
because it is a singlet under $SU(2)_R$. 
(If there are two Higgs multiplets in the bulk, one can also 
use $SU(2)_H$ flavor
symmetry to impose boundary conditions \cite{Barbieri:2001yz}).

The  squarks (also the Higgs scalars if they are
bulk fields) with a continuous Scherk-Schwarz 
phase have the mode expansion given by
\begin{eqnarray}\label{modeexpansion}
\left(\begin{array}{l}\phi \\
\phi^{c\dagger} \end{array}\right)(x,y)
=\frac{1}{\sqrt{2\pi R}}\sum_{n=-\infty}^\infty u_n(y)\,\varphi_n(x) 
\end{eqnarray}
where
\begin{eqnarray}
u_n=e^{-i\omega\sigma_2 y/R}\left(\begin{array}{l}\cos(ny/R) \\
\sin(ny/R) \end{array}\right)
\end{eqnarray}
and $(\Box-M^2_n)\varphi_n(x)=0$ with $M^2_n=(n+\omega)^2/R^2$.
Here we note that the orthogonality is defined for eigenstates of $SU(2)_R$ 
doublet as
\begin{eqnarray}
\frac{1}{2\pi R}\int^{2\pi R}_0 dy\, (u_n(y))^\dagger u_m(y)=\delta_{nm}.
\end{eqnarray}

\noindent
The spectrum of the bulk fields is
\begin{eqnarray}
m_{\psi_M,k} & = & \frac{k}{R}, \quad k\geq 0,
\qquad\qquad  m_{\psi^c_M,k}=\frac{k}{R}, \quad k\geq 1, \\
m_{\phi_M,k} &  = & m_{\phi^c_M,k}= 
\frac{k+\omega}{R}, \qquad\quad k\in {\bf Z}, \qquad
 (M\equiv Q, U).\\[7pt]
m_{\psi_H,k} & = & m_{\psi^c_H,k}=\frac{k+1/2}{R}, \qquad k\geq 0,
\\[7pt]
m_{\phi_H,k} & = & m_{\phi^c_H,k}=\frac{k+\omega}{R}, \qquad\quad
 k\in {\bf Z},
\end{eqnarray}
and $\eta_k^F=\eta_k^\phi=1/\sqrt{2}$ and $\eta_k^\psi=1/{\sqrt
  2}^{\delta_{k,0}}$. Finally, for gauginos one has that
  $m_{\lambda_{1,2},k}=(k+\omega)/R$, $(k\in\bZ)$.
We note that the zero mode of a bulk Higgs scalar acquires
a tree-level mass of $\omega/R$. 
In this case,  the one-loop Yukawa correction must be larger
than this tree-level mass for electroweak symmetry 
breaking \cite{Barbieri:2001dm}.
However, if the Higgs multiplet is a brane field, such situation is 
avoided, and  one can  still assume that there is no 
tree-level Higgs mass. 

The action is in this case similar to that in eq.(\ref{action4d})
except that the sums over $k$ and $l$ should be taken over the whole
set of  integer numbers. Then the one-loop mass corrections are
\begin{eqnarray}\label{mass2a}
-i\, m_{\phi_H}^2 (q^2)\bigg\vert_B\!\!\!\! \!&=& \!\!\!\!
- i \,  f_t^2 N_c\!\!\! \sum_{k\in {\bf Z},\, l\in {\bf Z}}
    \Big[\eta^{F_Q}_k \eta^{\phi_U}_l\Big]^2
   \!\! \int \frac{d^d p}{(2\pi)^4}
    \frac{2\,(p+q)^2 \,\,
\mu^{4-d}}{((p+q)^2+m_{\phi_Q^c,k}^2)(p^2+m_{\phi_U,l}^2)},
\nonumber\\[12pt]
-i \, m_{\phi_H}^2 (q^2)\bigg\vert_F\!\!\!\!\!
 &= &\!\!\!\!  i  \,f_t^2 N_c \!\sum_{k\geq 0,\, l\geq 0}
    \Big[\eta^{\psi_Q}_k \eta^{\psi_U}_l\Big]^2
    \!\int \frac{d^d p}{(2\pi)^4}
    \frac{2\,p.(p+q)\,\,\mu^{4-d}}{
((p+q)^2+m_{\psi_Q,k}^2)(p^2+m_{\psi_U,l}^2)}.
\\ \nonumber
\end{eqnarray}
Adding the bosonic and fermionic contributions, 
one obtains the one-loop mass correction as
\begin{eqnarray}\label{diva}
-m_{\phi_{H}}^2 (q^2) & = &
\frac{f_t^2}{16 \pi^3 R^2}N_c \int_0^1 dx\,
\Big[ \cJ_2[0,0,c]-  \cJ_2[\omega,\omega,c] \Big]
\nonumber\\
\nonumber\\
&+& \frac{f_t^2}{32\, \pi^2 } \,\,\kappa_\epsilon\,
N_c\,q^2 \int_0^1  \! dx  \Big[
x(x - 1) \cJ_1[0,0,c]
-(1-\!x)^2 \cJ_1[\omega,\omega,c] \Big].
\end{eqnarray}
Therefore we find again
\begin{eqnarray}
m_{\phi_{H}}^2(q^2)=m_{\phi_{H}}^2(0)-
\frac{f_t^2}{2^{8}}N_c\, q^4\,R^2\,
\bigg[\frac{1}{\epsilon}+\ln(2\pi R\,\mu)\bigg]
+  \frac{1}{R^2}\, \cO(q^2 R^2)
\end{eqnarray}
where $\cO(q^2 R^2)$ terms are  $\epsilon$-independent,
are due to the  two $\cJ_1$ functions in (\ref{diva})
and can be evaluated numerically using eqs.(\ref{a2}), (\ref{a3}). 
Higher derivative counterterms are again required,
and the same arguments as in eq.(\ref{hdo1}), (\ref{hdo2}) apply.

In this section we considered so far  Yukawa corrections only. 
Regarding the gauge correction, in both cases of 
discrete and continuous Scherk-Schwarz
twists, the spectrum and brane coupling
 of gaugino are the same as in the local supersymmetry  breaking.
 Hence  the resulting one-loop correction to a brane scalar mass 
is the same \cite{choi-lee} as before, eqs.~(\ref{eq26}) and (\ref{eq27}).

\section{Further remarks on higher derivative counterterms.}
 \label{section3}

In this section we discuss further the origin of the higher derivative 
counterterms  
and their relation to the local and non-local character of
supersymmetry breaking   of
Sections~\ref{section2.1},~\ref{section2.2}.

Let us recall first that 
the   radiative correction to the scalar mass from the
Kaluza-Klein modes has the general structure given by (\ref{thr})
below.  This equation is just a generalisation of eqs.(\ref{mass})
(\ref{mass1extra}), (\ref{mass2}), (\ref{mass2a}) for $m_{\phi_H}$,
 all recovered for particular values of
wavefunction coefficients  $\sigma_{1,2}=1,2$ and of
 ``mass shifts'' $c_{1,2}$.  After a long calculation  one 
obtains \footnote{To  derive 
eq.(\ref{thr}) one can use the method outlined in Appendix C of  
ref.\cite{Ghilencea:2004sq}.} ($d=4-\epsilon$)

\begin{eqnarray}\label{thr}
\cE(q^2)\!\!\! &\equiv  & \!\!\!\!\!\!\!\!
\sum_{k_1\geq 0; \, k_2\geq 0}
\bigg[\frac{\sigma_1}{2}\bigg]^{\delta_{k_1,0}}
\bigg[\frac{\sigma_2}{2}\bigg]^{\delta_{k_2,0}}
\!\!\int \frac{d^d p}{(2\pi)^d}
\frac{\alpha \, p^2+\beta\, (p.q)+\gamma \, q^2+\delta}{ 
[ (q+p)^2+(k_2+c_2)^2/R^2\, ] \,\, [\,p^2+ (k_1+c_1)^2\, u^2/R^2 ]}
\nonumber\\[9pt]
&=&\!\!\!\frac{1}{(4 \pi)^2}\,\frac{2}{\epsilon}\,
\bigg\{\,\,\Big(\delta+q^2(\gamma-\beta/2)\Big) 
\Big(c_1-\frac{\sigma_1-1}{2}\Big)
\Big(c_2-\frac{\sigma_2-1}{2}\Big)
\nonumber\\[9pt]
&&-\,\frac{\alpha\,u}{6\, R^2} \,
\bigg[u\,  c_1 \,\Big(c_2-\frac{\sigma_2 -1}{2}\Big)
\Big(1+3 c_1 (1-\sigma_1) +2 c_1^2\Big)
+ (c_1\!\leftrightarrow\!  c_2,\, \sigma_1\!\leftrightarrow\! \sigma_2,
\, u \!\rightarrow\! \frac{1}{u})\bigg]
\nonumber\\[9pt]
&&- 
\,\bigg[\frac{\pi^2}{32 u}\,\, \delta\, q^2 R^2 + \frac{\pi^2}{2^8 u}
  \,\,(\alpha-4\beta+8\gamma)\, q^4 R^2\bigg]\,\,\bigg\}+\cO(\epsilon^0).
\end{eqnarray}
This  result  is computed in the DR scheme and 
as usual, $\epsilon$ is a regulator of both the integral
{\it and} the double series in front of it; $\epsilon$ is thus  a
genuine 6D regulator rather than a 4D one\footnote{If the series 
in eq.(\ref{thr}) were restricted to a finite number of modes then 
$\epsilon$ would act as a 4D regulator only.}. One firstly performs the
momentum integral to obtain  a double series
whose summand (function of $k_1, k_2$) has powers involving $\epsilon$
and is multiplied by Gamma functions of $\epsilon$-dependent 
 argument. The series are analytically continued and their leading  
contribution to  $\cO(\epsilon^0)$ is obtained. 
Finally one takes account of the Gamma functions and finds the above
result\footnote{In a similar way one can also find the finite,
$\cO(\epsilon^0)$ terms of (\ref{thr}).}. Note that the sums in 
(\ref{thr}) are restricted to positive integers only, unlike 
the expressions of  $\cJ_{1,2}$ used previously and
which involve double sums over $\bZ$. 
The motivation and the advantage of using eq.(\ref{thr}) is that (unlike 
the analysis using $J_{1,2}$) it will allow
us to see explicitly  how quadratic divergences cancel.

The divergences in (\ref{thr}) are: $q^2/\epsilon$ which
account for wave function renormalisation, the terms
 $1/(R^2 \epsilon)$ which account for
quadratic divergences and finally, the
terms $q^4 R^2/\epsilon$ which account for quartic 
divergences (higher derivative counterterms). 

We can now apply the result 
(\ref{thr}) to the calculations in 
 Sections 2.1 and  2.2 to discuss the origin of the higher 
derivative operators and of other divergences  present.
For the case in Section~\ref{section2.1.1}
 for  the contribution in the first line
of eqs.(\ref{mass}), (\ref{mass1extra}) one has
\begin{eqnarray}\label{prt0}
\rm{Bosonic \,\,\, part:} &\qquad\,&
\sigma_1=2, \,\sigma_2=1,\, \beta=2,\, 
\gamma=1,\,c_1=1/2, \,c_2=0. 
\nonumber\\
\rm{Fermionic\,\,\, part:}&\qquad\,&
\sigma_1=1,\, \sigma_2=1,\, \beta=1, \,\gamma=0,\,
 c_1=0,\,\,\,\,\,\,\,c_2=0.
\end{eqnarray}
while for the case in Section \ref{2.2.1}, 
eq.(\ref{mass2}) one has  
\begin{eqnarray}\label{prt}
\rm{Bosonic \,\,\, part:} &\qquad\,&
\sigma_1=2, \,\sigma_2=2,\, \beta=2,\, 
\gamma=1,\,c_1=1/2,\,c_2=1/2. 
\nonumber\\
\rm{Fermionic\,\,\, part:}&\qquad\,&
\sigma_1=1,\, \sigma_2=1,\, \beta=1, \,\gamma=0,\,
 c_1=0,\quad c_2=0.
\end{eqnarray}
In both cases $\alpha=1$, $u=1$, $\delta=0$. 

Finally, for the case in Section \ref{section2.2.2} 
with a continuous Scherk-Schwarz phase,
one has the same expression for the fermionic part as in (\ref{prt}), 
but the bosonic part is given by
\begin{eqnarray}\label{prt3}
&&\frac{1}{4}\,\bigg[\,\cE\Big(q^2;c_1=c_2=\omega\Big)
+\cE\Big(q^2;c_1=\omega,c_2=1-\omega\Big) \nonumber \\
&&\,\, +\,\cE\Big(q^2;c_1=1-\omega,c_2=\omega\Big)
+\cE\Big(q^2;c_1=c_2=1-\omega\Big)\,\bigg]
\end{eqnarray}
where $\sigma_1\!=\!\sigma_2\!=\!2,\beta\!=\!2,\gamma\!=\!1$ 
and $\alpha\!=\!1,u\!=\!1,\delta\!=\!0$
in each term. Each contribution $\cE$ in (\ref{prt3})
has quadratic divergences $1/(R^2 \epsilon)$, as seen from 
eq.(\ref{thr}).
However they cancel in the sum of  the first two terms (of arguments 
$c_2=\omega$ and $c_2=1-\omega$) and also in the sum of the last two 
terms. This is essentially due to summing over the whole set 
$\bZ$ of modes in  eq.(\ref{modeexpansion}). 
Given the above values of $c_{1,2}, \sigma_{1,2}$, 
for all cases considered in eq.(\ref{prt0}), (\ref{prt}), (\ref{prt3}),
one concludes that
 the coefficient of the quadratic
divergences $1/(R^2 \,\epsilon)$ vanishes separately  for the bosons
and fermions. Note that the values of  $c_i=0, 1/2$ are special for in
that case (with corresponding values of $\sigma_i$), $\cE$ alone 
has no quadratic divergence. 

For the origin of higher derivative operators it
 is important to notice
 that the coefficient of  $q^4 R^2/\epsilon$ term
 is independent of $c_{1,2}$ and also of $\sigma_{1,2}$. Note that
$c_{1,2}$ which enter the mass formulae for the Kaluza-Klein states
are in fact set by  the boundary conditions for the hypermultiplets 
with respect to  the compact dimension.
Therefore, this coefficient is independent of the phase that
hypermultiplet fields have with respect to this  dimension
and, to some extent\footnote{see discussion in the last paragraph of
 this section.}, 
on the way supersymmetry is broken.
The coefficient of $q^4 R^2/\epsilon$
depends only on $\alpha,\,$ $\beta,\,$ $\gamma$, which in 
turn are  controlled by the nature (fermionic/bosonic) of the component
fields (via their propagator in momentum space). 
 Technically, the term $q^4 R^2/\epsilon$
is strictly the result of the presence  of {\it two} 
sums in front of the integral (\ref{thr}), over terms with
$k_i\not=0,\, i=1,2.$ (this explains the 
absence of such divergences and corresponding 
 higher dimensional counterterms
 in the 4D theory\footnote{This is unlike the
 case of $1/(R^2\epsilon)$ divergences 
 to which all modes including $k_i=0$
 contribute.} of {\it zero-modes, $k_i=0$}).  Therefore,
divergent terms $q^4 R^2/\epsilon$  can be avoided at one-loop 
provided that there is 
only one bulk propagator (see Table 1 for details when this can happen).

With these considerations one concludes that the 
presence of $q^4 R^2/\epsilon$
and thus of the higher derivative counterterms is related 
to the multiplicity of the modes.
Such operators are then due to
 the non-renormalisability of the models - 
initial supersymmetry cannot protect against their emergence 
as counterterms, regardless of the way supersymmetry is broken.  
Our analysis also shows that such operators 
are most relevant in models with low compactification scales,
when their effects are less suppressed, see 
eqs.(\ref{hdo0}), (\ref{hdo2}).

Finally, note  the presence of a  $u$ dependence of the coefficient
of  $q^4 R^2/\epsilon$  in (\ref{thr}), 
which is present in  the case of  brane-localised
supersymmetry breaking (see Section 2.1). This signals
some dependence between this supersymmetry breaking 
mechanism and the coefficient of the higher derivative 
operators. No such dependence appears for the discrete or
continuous Scherk-Schwarz mechanism.

\section{Phenomenological implications: living with ghosts?}\label{section4}

The presence of higher derivative operators in the action of the
 scalar field $\phi_H$ has implications for phenomenology. Their
investigation is however  difficult since  
theories with higher derivative operators can bring in    
more fundamental problems, such as unitarity violation, 
non-locality effects, the presence of additional ghost fields, and for
 these reasons such theories were less popular in the past 
(for some studies of such theories see for example
 \cite{Pais}-\cite{Mannheim:2004qz}).
Therefore the phenomenological considerations below
should be taken with due care. 

In the presence of higher derivative operators, 
 new fields (ghosts) are present. 
To see this, let us write the propagator of  $\phi_H$
in the presence of the higher derivative operator found in
eqs.(\ref{hdo0}), (\ref{hdo1}), (\ref{hdo2}):
$\cL_4= -\xi\, R^2 \,\phi_H \Box^2 \phi_H+\cdots$, where $\xi$
is an arbitrary constant (assumed positive for the convergence of
 the partition function); if $\phi_H$
is a bulk field,  we refer to its zero-mode only.
The  propagator of $\phi_H$  changes then  into

\begin{eqnarray}\label{propagator}
\frac{1}{-\xi R^2\, p^4+p^2-m^2}=\frac{1}{(1-4 \,\xi\, R^2 \, m^2)^{\frac{1}{2}}}\,
\bigg[\frac{1}{p^2-m_-^2}-\frac{1}{p^2-m_+^2}\bigg]
\end{eqnarray}
with 
\begin{eqnarray}
m_{\pm}^2=\frac{1}{2 \,\xi R^2}\Big[1\pm (1-4 \,\xi\, R^2\,m^2)^{\frac{1}{2}}\Big]
\end{eqnarray}
The second term in the rhs of (\ref{propagator}) has the ``wrong''
sign, thus it signals the presence in the model of a 
ghost field of mass $m_+$. Here  $m^2$ is the  one-loop induced
mass $m_{\phi_H}^2(0)$  of $\phi_H$ plus the tree level contribution (if any),  
and $m_-^2$ is its value  corrected by the higher
derivative operator, but ignoring  loop corrections $\cO(q^2
R^2)$ of eq.(\ref{one-loop_c2}) at $q^2\!=\!m_-^2$. In function of the
supersymmetry breaking mechanism which controls the coefficients
$c_{1,2}$ in (\ref{j12functions}), one may have $m^2_{\phi_H}(0)<0$
and thus  electroweak symmetry breaking (assuming no tree level mass is present).
With $m^2<0$,  $\xi>0$ then $m^2_+>0$ and $m^2_-<0$, and the 
symmetry breaking may be maintained.

In the following we require that $m_+^2>M_*^2$, i.e. 
the ghost mass $m_+$ is larger than the cutoff
$M_*$ of our 5D theory, which for an effective theory approach as ours
seems a  natural requirement. We then study its implications  for
$m_-^2$. For this purpose we  use  a  dimensional analysis 
\cite{Chacko:1999hg,Buchmuller:2005rt}
 to obtain   perturbativity constraints on $M_*$, eqs.(\ref{bound_g}),
(\ref{bound_y}), by requiring the effective
gauge and Yukawa couplings be less than unity. The result is
\begin{eqnarray}\label{bnd}
M_* \!< \!\frac{12 \pi^2}{\delta}\,\frac{1}{g_4^2 R}, \qquad
M_* \!< \!\frac{12\pi^2}{\delta}\,\bigg[\frac{\delta}{16
    \,\pi^2\,f_{4,t}^{2/(p-1)}}\bigg]^{\frac{p-1}{p}}
 \frac{1}{R}.
\end{eqnarray}
The first (second) bound is from gauge (Yukawa) interaction. Here
 $p=2$ ($p=3$) if the Higgs field is a brane (bulk)
field;  $\delta$ is a factor equal to N for SU(N) gauge group,
taking account of number of degrees of freedom present in loops 
\cite{Chacko:1999hg}. 
Further, one imposes the first condition in the equation below,   
solves it for $\xi$ which is then used to evaluate  the change in $m_-^2$, to find:
\begin{eqnarray}
m_+^2=\gamma \,M_{*, u}^2,\qquad\Rightarrow \qquad \xi=\frac{\gamma \, \nu^2
  -\beta\, {\it sgn}\,(m^2)}{\gamma^2\, \nu^4},\qquad
\Delta_-=\frac{1}{{\it sgn}\,(m^2)\,\gamma
  \,\nu^2/\beta-1}
\end{eqnarray}
Here $M_{*, u}$ is  upper bound  of $M_*$ in (\ref{bnd}),
 $\nu$ is the factor multiplying $1/R$ in $M_{*, u}$, 
 $\beta=\vert m^2\vert R^2$, {\it sgn(x)} is $+1\, (-1)$ for
 $x\!>\!0$\, ($x\!<\!0$),
 and $\Delta_-$ is the 
 variation of $m_-^2$ relative to $m^2$. 
One should then take
 $\gamma>1$ but $\gamma\sim \cO(1)$ may still satisfy $m_+^2>M_*^2$. 
We consider now  the Yukawa correction only and 
take\footnote{These correspond
  to the model with discrete Scherk-Schwarz supersymmetry breaking
 discussed above.}  $\delta=3$ for SU(3)$_c$, $m^2<0$ and $f_{4,t}\simeq 1$ for 
which  $\beta \sim 0.1$.
This gives for $\gamma=1$,  $\xi=0.03 (0.12)$ and 
$\Delta_-=-0.3\%\,(-1.2\%)$ for  $p=2\,(3)$. 
Therefore the coefficient of the higher derivative
counterterm must be very small and its correction 
to the scalar field mass is  negligible. A value $\gamma=1/4$
 would give   $\xi=0.13 (0.53)$, 
$\Delta_-=-1.3\%\,(-4.8\%)$ for $p=2\,\,(3)$, thus  $\Delta_-$  is mildly changed.
Note that this discussion ignored  the additional terms ($\cO(q^2
 R^2)$) at  $q^2=m_-^2$. For the implications on the
{\it physical} mass of the scalar field and its dependence on the parameter
$\xi$ one must actually evaluate the minimum of the
one-loop potential computed in the presence of such operators, which
is beyond the purpose of this work. 
Finally, if the condition $m_+^2>M_*^2$  is not satisfied, 
 the presence of
the ghost pole in the effective theory with the cutoff scale $M_*$
requires further theoretical investigation\footnote{for a discussion see 
\cite{Smilga:2005gb}.}.

\section{Conclusions.}

In this work we addressed the role that higher derivative operators
play in the study of radiative corrections to the mass of the 
(Higgs) scalar field in 
5D N=1 supersymmetric models compactified on $S_1/Z_2$.
 This is an important issue
because it addresses  how physics associated with compact dimensions 
decouples at low energies $q^2\ll 1/R^2$,
how these operators control the running 
of the scalar field  mass across $q^2\sim 1/R^2$,
and the UV behaviour of the mass under the scaling of the
momenta to $q^2\gg 1/R^2$. Our technical results can be 
applied to a large class of orbifold models which investigate
such one-loop effects from the bulk fields.

The interactions considered were localised
superpotentials and gauge interactions. Although the models
are rather minimal, the multiplet structure and the
interactions considered  are generic  in any realistic 
higher dimensional extensions of the SM or its supersymmetric
versions. It was found that Yukawa interactions, unlike 
gauge corrections, can generate higher derivative counterterms to the
scalar field mass even at the one-loop level.

The work  examined closely the relationship of the
higher derivative  operators with the way
supersymmetry breaking was transmitted to the visible sector.
The supersymmetry breaking scenarios addressed were local
supersymmetry breaking by the F-term of a gauge singlet  field 
localised at a hidden brane (fixed point) and 
also non-local breaking via Scherk-Schwarz (discrete/continuous) 
boundary conditions.
In the  case of local  breaking, this effect  
is transmitted via radiative corrections to the
 scalar field,  associated with the
Higgs field. Such corrections, due to the compact dimension, are 
induced  by the bulk fields which feel the supersymmetry breaking at 
the hidden brane. The second case, of  non-local supersymmetry
breaking, considered  supersymmetry breaking by discrete and
continuous  Scherk-Schwarz twists of the bulk fields by using the 
$SU(2)_R$ symmetry of the multiplet 
content.

Our analysis showed that  in all these cases of supersymmetry breaking
the emergence of higher derivative counterterms to the mass of the
scalar field  is very similar
and has little or no dependence on the details of the breaking
mechanism considered.
This is the main result of the paper.
At the technical level this means 
 that the coefficient of these operators is independent of the
boundary conditions for the bulk fields. As a result 
these operators seem to be a generic presence in orbifold
compactifications and initial supersymmetry cannot protect against their
presence, in some cases even at the one-loop level. This result is
ultimately due to the non-renormalisable character of
the theory and may also raise questions on the power of
initial 5D supersymmetry  in maintaining a mild UV behaviour 
in the compactified theory.

The phenomenological implications of the presence
of higher derivative operators in the action of the scalar
field were  briefly investigated. It was found that the  requirement
that the 5D effective theory be weakly coupled together with the 
ghost field mass be larger than the 5D effective cutoff 
lead  to  small corrections on the scalar field mass in the action.
This can change dramatically  when any of these constraints is
relaxed.  To evaluate  the {\it physical} mass one is required 
to compute the one-loop corrected scalar potential {\it in the presence} 
in the action of  higher derivative operators. Such calculation requires however 
a prior and  more comprehensive  study of the theories  with higher derivative
 operators  at the quantum  level, which is beyond our purpose.

Our  study is also important because other approaches to
compactification like string theory currently shed little light on such
issues. The reason for this is that in string theory
one computes the scalar potential (vacuum energy),
derived at zero external
momenta. Its second derivative giving a scalar mass
cannot then recover a momentum dependence (``running'') of the latter. We are 
thus confined to study these higher derivative operators 
in the framework of field theory  orbifolds. The situation is very
similar to the case of one-loop corrections to the gauge couplings, 
where again the role of higher derivative operators  cannot be 
discussed  in 
(on-shell) string loop corrections, but  can be evaluated consistently 
in the context of  field theory orbifolds 
\cite{Ghilencea:2003xj,Nibbelink:2005vi,Kazakov:2002jd}.

\section*{Acknowledgements.}
The work of D. Ghilencea was supported by a post-doctoral research
grant from the Particle Physics and Astronomy Research Council
(PPARC) United Kingdom.

\newpage
\section*{Appendix}
\label{appendix}
\def\theequation{\thesubsection-\arabic{equation}} 
\def\thesubsection{A} 
\setcounter{equation}{0}
\subsection{Series of Integrals in the DR scheme.}\label{appendixA}

The functions $\cJ_{1,2}$ 
used in  
eqs.(\ref{j12functions}), (\ref{mass3}), (\ref{div}), (\ref{diva}) 
 are, up to  a re-definition of $\epsilon$, 

\begin{eqnarray}\label{j1general}
\!\!
\cJ_p[c_1,c_2,c] 
 \equiv   \!\! \sum_{n_1,n_2\in\bZ}
\int_0^\infty \frac{dt}{t^{p+\epsilon}}
\, e^{-\pi \, t\, [\,c+a_1 (n_1+c_1)^2+a_2 (n_2+c_2)^2]},\quad  
a_{1,2}>0,\, c\geq 0; \,\, p=1,2.
\end{eqnarray}

\noindent
The expressions given below for  $\cJ_p$, $p=1,2$  generalise 
results quoted in  Appendix B of  \cite{Ghilencea:2004sq} 
valid only  for the case $c/a_{1,2}\leq 1$ and which are 
included  here for completeness. Using  the method in Appendix B of 
\cite{Ghilencea:2003xy} one has that, if $0\leq c/a_1<1$

\begin{eqnarray}\label{a2}
\!\!\cJ_1[c_1,c_2,c] \!
& =&  \!\!\! \frac{\pi c}{\sqrt{a_1 a_2}} \frac{1}{\epsilon}+
\frac{\pi c}{\sqrt{a_1 a_2}} 
\ln\Big[4\pi \,a_1\,
    e^{\gamma+\psi(\Delta_{c_1})+\psi(-\Delta_{c_1})}\Big] 
\nonumber\\
\nonumber\\
&+& 2\pi u\,
\Big[\frac{1}{6}+\Delta_{c_1}^2-\Big(c/a_1+\Delta_{c_1}^2\Big)^\frac{1}{2}\Big]
 -   \sum_{n_1\in\bZ}
\ln\Big\vert 1-e^{-2\pi \,\gamma (n_1)}\Big\vert^2
\nonumber\\
\nonumber\\
& +\!&\!\! \sqrt\pi\, u\,
\!\sum_{p\geq 1}^\infty \frac{\Gamma[p\!+\!1/2]}{(p\!+\!1)!}
\bigg[\frac{-c}{a_1}\bigg]^{p+1} 
\!\!
\Big(\zeta[2p\!+\!1,1\!+\! \Delta_{c_1}]\!+\!
\zeta[2p\!+\!1,1\!-\!\Delta_{c_1}]\Big),
\\
\nonumber
\end{eqnarray}
with $u\!\equiv\!\sqrt{a_1/a_2}$, $\gamma\!=\!0.577216..$.
 If $c/a_1\!\geq\! 1$, one has
the same pole in $\epsilon$, but different finite part:

\begin{eqnarray}\label{a3}
\cJ_1[c_1,c_2,c] 
&=&  \!\!\! \frac{\pi c}{\sqrt{a_1 a_2}} \frac{1}{\epsilon}+
\frac{\pi c}{\sqrt{a_1 a_2}} \ln\Big[4\pi \,c \,
    e^{\gamma-1}\Big]  -\!\sum_{n_1\in\bZ}\ln\Big\vert
 1\!-\!e^{-2\pi \,\gamma (n_1)}\Big\vert^2\,
\nonumber\\
\nonumber\\
\!&\!+&\!  4 \bigg[\frac{c}{a_2}\bigg]^{\frac{1}{2}}\!\!
\sum_{\tilde n_1>0}
\frac{1}{\tilde n_1}\,\, 
{\cos(2\pi \tilde n_1\,c_1)} \,\,K_1\Big(2\pi\tilde n_1
(c/a_1)^{\frac{1}{2}}\Big),\qquad\qquad\qquad\qquad
\\
\nonumber
\end{eqnarray}
with the notation
\begin{eqnarray}\label{notation}
\gamma(n_1)&=&\frac{1}{\sqrt{a_2}}
\,\big[z(n_1)\big]^{\frac{1}{2}}-i\,c_2,
\nonumber\\
z(n_1)&=& c + a_1 (n_1+c_1)^2.
\end{eqnarray}
Here     $\zeta[z,a]$ is the Hurwitz  
Zeta function, $\zeta[z,a]=\sum_{n\geq 0} (n+a)^{-z}$, $\rm{Re}\,
z>1$,\,$a\not=0,-1,-2,\cdots$, and  $\psi(x)=d/dx \,\ln \Gamma[x]$. 
Eqs.(\ref{a2}), (\ref{a3}) depend on the fractional part of
$c_{1,2}\,$ defined by
$\Delta_{c_i} \equiv c_i-[c_i]$ with $0\!\leq\!
\Delta_{c_i}\!<\! 1$,
$[c_i]\in \bZ$. Finally, $K_n$ is the modified Bessel function
\cite{gr} 

\begin{equation}\label{bessel1}
\int_{0}^{\infty} \! dx\, x^{\nu-1} e^{- b x^p- a
x^{-p}}=\frac{2}{p}\, \bigg[\frac{a}{b}
\bigg]^{\frac{\nu}{2 p}} K_{\frac{\nu}{p}}(2 \sqrt{a \, b}),\quad Re
(b),\, Re (a)>0
\end{equation}
with
\begin{eqnarray}
K_1[x]=e^{-x}\sqrt{\frac{\pi}{2
    x}}\left[1+\frac{3}{8 x}-\frac{15}{128 x^2}
+\cO(1/x^3)\right]
\\
\nonumber
\end{eqnarray}
which is  strongly suppressed at large argument.

One also finds that, if $c\ll 1$

\begin{eqnarray}
\cJ_1[c_1,c_2,c\ll 1] 
& =&  \frac{\pi c}{\sqrt{a_1 a_2}} \frac{1}{\epsilon}
-\ln\bigg\vert\frac{\vartheta_1(c_2-i u c_1\vert i u)}{(c_2-i u
  c_1)\,
\eta(i u) } \,e^{-\pi u c_1^2}\bigg\vert^2
-\ln(c/a_2+\vert c_2-i u c_1\vert^2)
\nonumber\\
\nonumber\\
\cJ_1[1/2,1/2,c\ll 1] 
& =&  \frac{\pi c}{\sqrt{a_1 a_2}} \frac{1}{\epsilon}
-\ln\bigg\vert\frac{\vartheta_1(1/2-i u/2\vert i u)}{\eta(i\, u) } 
\,e^{-\pi u/4}\bigg\vert^2
\nonumber\\
\nonumber\\
\cJ_1[1/2,0,c\ll 1]&=&\frac{\pi c}{\sqrt{a_1 a_2}} \frac{1}{\epsilon}
-\ln\bigg\vert\frac{\vartheta_1(-i u/2\vert i u)}{(u/2)\,
\eta(i u) } \,e^{-\pi u/4}\bigg\vert^2
-\ln[ u/2]^2
\nonumber\\
\nonumber\\
\cJ_1[0,0,c\ll 1] 
 &=&  \frac{\pi c}{\sqrt{a_1 a_2}} \frac{1}{\epsilon}
-\ln\Big[ \, 4\pi^2 \, \vert\eta(i \, u) \vert^4 
\,\,a_2^{-1}\Big]-\ln c
,\qquad u\equiv \sqrt{a_1/a_2}\\
\nonumber
\end{eqnarray}
used in the text. Above we used  the Dedekind Eta function
\begin{eqnarray}\label{ddt}
\eta(\tau) & \equiv & e^{\pi i \tau/12} \prod_{n\geq 1} (1- e^{2 i
\pi\tau\, n}),
\nonumber\\
\eta(-1/\tau) & = & \sqrt{-i \, \tau}\,\eta(\tau),
\qquad
\eta(\tau+1)=e^{i\pi/12}\eta(\tau),
\end{eqnarray}
and the Jacobi Theta function  $\vartheta_1$ \cite{gr}

\begin{eqnarray}\label{th}
\vartheta_1(z\vert\tau)&\equiv & 2 q^{1/8}\sin (\pi z) \prod_{n\geq 1} 
(1- q^n) (1-q^n e^{2 i \pi z}) (1- q^n e^{-2 i \pi z}), \qquad 
q\equiv e^{2 i \pi \tau}
\nonumber\\
 \nonumber\\
  &=& - {i}\sum_{n\in\bZ} (-1)^n e^{i\pi\tau (n+1/2)^2}\,
  e^{(2n+1)i\pi z} 
\end{eqnarray}
which has the properties 
\begin{eqnarray}
\vartheta_1(\nu\,|\,\tau+1) & =& e^{i\pi/4} \, \vartheta_1(\nu|\tau),
\nonumber\\
\vartheta_1(\nu+1\,|\,\tau)& = & -\, \vartheta_1(\nu|\tau),
\nonumber\\
\vartheta_1(\nu+\tau\,|\,\tau)& = & -\,   e^{-i\pi \tau -2 i \pi \nu} \, 
\vartheta_1(\nu|\tau)
\nonumber\\
\vartheta_1(-\nu/\tau\, |-1/\tau)& =& 
e^{i\pi/4} \tau^{1/2} \exp(i\pi\nu^2/\tau)
\, \,\vartheta_1(\nu|\tau)
\label{R3_3}
\end{eqnarray}
In the following we provide the results for
 $\cJ_2[c_1,c_2,c]$ whose calculation is almost identical.

\noindent
If $0\leq c/a_1<1$
\begin{eqnarray}\label{pp1}
\!\!\cJ_2[c_1,c_2,c] \!
& =&  \!\!\! -\frac{\pi^2 c^2}{2\, \sqrt{a_1 a_2}} \frac{1}{\epsilon}
-\frac{\pi^2 c^2}{2 \,\sqrt{a_1 a_2}} \ln\Big[4\pi \,a_1\,
    e^{\gamma+\psi(\Delta_{c_1})+\psi(-\Delta_{c_1})}\Big] 
\nonumber\\
\nonumber\\
&+& \pi^2 a_2 u^3 \,\frac{1}{3}
\bigg\{
\frac{1}{15}-2\Delta_{c_1}^2 (1+ \Delta_{c_1}^2)
  -6\frac{c}{a_1}\Big[\frac{1}{6}+\Delta_{c_1}^2\Big]
+4 \Big[c/a_1+\Delta_{c_1}^2\Big]^\frac{3}{2}\bigg\}
\nonumber\\
\nonumber\\
&+&
\sum_{n_1\in\bZ}\left\{ \big(a_2 \,z(n_1)\big)^{\frac{1}{2}}\,\,
\Li_2 ( e^{-2\pi\gamma(n_1)})+\frac{a_2}{2\pi}\,\,
\Li_3 ( e^{-2\pi\gamma(n_1)}) +c.c.\right\}
\nonumber\\
\nonumber\\
&+&
\!\! \pi^{3/2} a_2 \, u^3\, 
\sum_{p\geq 1}^\infty \frac{\Gamma[p\!+\!1/2]}{(p\!+\!2)!}
\bigg[\frac{-c}{a_1}\bigg]^{p+2} 
\!\!
\Big(\zeta[2p\!+\!1,1\!+\! \Delta_{c_1}]\!+\!
\zeta[2p\!+\!1,1\!-\!\Delta_{c_1}]\Big).
\\
\nonumber
\end{eqnarray}

\noindent
If  $c/a_1\geq 1$  the pole structure is similar:

\begin{eqnarray}
\!\!\cJ_2[c_1,c_2,c] \!
& =&  \!\!\! -\frac{\pi^2 c^2}{2 \,\sqrt{a_1 a_2}} \frac{1}{\epsilon}
-\frac{\pi^2 c^2}{2 \sqrt{a_1 a_2}} \ln\Big[\pi \,c\,\,
    e^{\gamma-3/2}\Big] 
\nonumber\\
\nonumber\\
&+&
\sum_{n_1\in\bZ}\left\{ \big(a_2 \,z(n_1)\big)^{\frac{1}{2}}\,\,
\Li_2 ( e^{-2\pi\gamma(n_1)})+\frac{a_2}{2\pi}\,\,
\Li_3 ( e^{-2\pi\gamma(n_1)}) +c.c.\right\}
\nonumber\\
\nonumber\\
&+&
4\,c \,u\,
 \sum_{\tilde n_1>0}\frac{1}{\tilde n_1^2}\,
\cos(2\pi\tilde n_1 c_1)\,K_2\Big(2\pi\,\tilde n_1
(c/a_1)^{1/2}\Big).
\\
\nonumber
\end{eqnarray}
For the simpler case $c \ll 1$,

\begin{eqnarray}
J_2[c_1,c_2,c\ll 1] &=&
-\frac{\pi^2 c^2}{2\, \sqrt{a_1 a_2}} \frac{1}{\epsilon}
+ \pi^2 a_2\, u^3\,\frac{1}{3}\,
\Big[
\frac{1}{15}-2\Delta_{c_1}^2 (1- \Delta_{c_1})^2\Big]
\nonumber\\
\nonumber\\
&+&
\sum_{n_1\in\bZ}\left\{\sqrt{a_1 a_2}\,\vert n_1+c_1\vert\, 
\Li_2 ( e^{-2 \pi\sigma_{n_1}})+\frac{a_2}{2\pi}\,\,
\Li_3 ( e^{-2  \pi\sigma_{n_1}}) +c.c.\right\}\qquad\quad
\end{eqnarray}
with 
$\sigma_{n_1}=i\, c_2+\,u \,\vert n_1+c_1\vert$.
Also

\begin{eqnarray}
\!\!\!
J_2[0,0,c\ll 1] &=&
-\frac{\pi^2 c^2}{2\, \sqrt{a_1 a_2}} \frac{1}{\epsilon}
+\frac{\pi^2}{45}\, a_2\, u^3 \,
\nonumber\\
\nonumber\\
&+&\!\!\! \frac{a_2}{\pi} \!\!
\sum_{n_1\in\bZ}\Big\{2\pi \,u\,\vert n_1\vert\, 
\Li_2 ( e^{-2 \pi\sigma_{n_1}})+
\Li_3 ( e^{-2  \pi\sigma_{n_1}})\Big\}\qquad\qquad
\end{eqnarray} 
with $\sigma_{n_1}=u\,\vert n_1\vert$. 
Finally
\begin{eqnarray}\label{pp2}
\!\!\!\!
J_2[1/2,1/2,c\ll 1] &=&
-\frac{\pi^2 c^2}{2\, \sqrt{a_1 a_2}} \frac{1}{\epsilon}
-\frac{7\pi^2}{360}\, a_2 \,u^3 \,
\nonumber\\
\nonumber\\
&+&\!\!\!\frac{a_2}{\pi} \!\!
\sum_{n_1\in\bZ}\Big\{ 2\pi \,u\,\vert n_1+1/2 \vert\, 
\Li_2 (- e^{-2 \pi\sigma_{n_1}})+
\Li_3 (- e^{-2  \pi\sigma_{n_1}})\Big\}\qquad
\\
\nonumber
\end{eqnarray}
with $\sigma_{n_1}=u\, \vert n_1+1/2\vert$.

\vspace{1.5cm}

\def\theequation{\thesubsection-\arabic{equation}} 
\def\thesubsection{B} 
\setcounter{equation}{0}
\subsection{Upper bounds on $M_*$ from Dimensional Analysis.}
\label{appendixB}

We derive  the  upper bounds on the cutoff $M_*$,  used in the text, 
eqs.(\ref{bnd}).
For this purpose,  we first outline  the general dimensional analysis performed in 
 \cite{Chacko:1999hg} (Section~3.2) and \cite{Buchmuller:2005rt}
 (Section~3).  We then apply it  to our case to 
derive upper bounds on $M_*$ from perturbativity constraints.

The action in $D$ dimensions (for orbifolds with $y_i$ fixed points)
has the  form 
\bea\label{Bq1}
{\cal L}={\cal L}_{\rm bulk}(\phi)
+\sum_i\delta^{D-4}(y-y_i)\,{\cal L}_i(\phi,\psi_i)
\eea
where $\phi$  ($\psi$) is a bulk (brane)  field, respectively.
One can  assume canonical kinetic terms in (\ref{Bq1}) and then 
rescale these fields to their dimensionless counterparts 
$\hat\phi$, $\hat\psi$ and also  the derivatives as
\bea\label{Bq2}
\phi(x,y)=\left(\frac{M_*^{D-2}}{l_D/\delta}\right)^{\frac{1}{2}}
\!\!{\hat\phi}(x,y),\qquad
\psi_i(x)=\left(\frac{M_*^2}{l_4/\delta}\right)^{\frac{1}{2}}\!\!{\hat\psi}_i(x), 
\qquad\partial=M_*{\hat\partial},\label{rescale}
\eea
where $M_*$ is the cutoff scale; $l_D$ is a suppression factor 
which accounts for angular integrations of loop corrections in D dimensions
$l_D=(4\pi)^{D/2}\Gamma(D/2)$ which grows rapidly with D, 
and $\delta$ is the multiplicity of fields in loop diagrams 
for non-Abelian groups,  for example $\delta=N$ for $SU(N)$, $\delta=8$ for $SO(10)$.
Using  eq.(\ref{Bq2}) in eq.(\ref{Bq1}) gives

\be
{\cal L}=\frac{M_*^D}{l_D/\delta}\, \,{\hat{\cal L}}_{\rm bulk}(\hat\phi)
+\sum_i\delta^{D-4}(y-y_i)\,\,\frac{M_*^4}{l_4/\delta}\,\,
{\hat{\cal L}}_i(\hat\phi,\hat\psi_i).
\ee 
where ${\hat{\cal L}}_{\rm bulk},{\hat{\cal L}}_i$ only contain dimensionless
couplings and fields. 
If all couplings in ${\hat{\cal L}}_{\rm bulk},{\hat{\cal L}}_i$ are
of order $1$, all loops are of the same order of magnitude. The theory with 
 ${\hat{\cal L}}_{\rm bulk},{\hat{\cal L}}_i$ remains weakly coupled
if these dimensionless effective couplings remain smaller than unity
\cite{Chacko:1999hg,Buchmuller:2005rt}.

Let us apply this result  to  (bulk) gauge interactions.
The relation between the $D$ dimensional gauge coupling 
and the 4D gauge coupling is
\be
\frac{V_{D-4}}{g^2_D}=\frac{1}{g^2_4}
\ee
where $V_{D-4}$ is the volume of extra dimensions.
From the rescaled covariant derivative,
\bea
{\hat D}_\mu&=&\frac{\partial_\mu}{M_*}-\frac{ig_DA_\mu}{M_*}
=\frac{\partial_\mu}{M_*}
-ig_D\left(\frac{M_*^{D-4}}{l_D/\delta}\right)^{\frac{1}{2}}{\hat A}_\mu,
\eea
we identify the dimensionless parameter corresponding to the gauge
coupling as 
\be
g_{\rm eff}^2\equiv g^2_D\frac{M_*^{D-4}}{l_D/\delta}<1.
\ee
The above condition for perturbativity imposes the bound on the cutoff 
as
\be
M_*<\bigg(\frac{l_D/\delta}{g^2_4 \,V_{D-4}}\bigg)^{\frac{1}{D-4}}
\label{gbound}
\ee
Let us now consider a  brane-localized interaction

\be
{\cal L}_F=\int d^2\theta \,W(\Phi,\Psi)+{\rm h.c.}
\ee
where $\Psi$ ($\Phi$)  is a brane ($Z_2$ even bulk) multiplet
respectively. Under the rescaling 

\bea
\Phi(x,y)=\left(\frac{M_*^{D-2}}{l_D/\delta}\right)^{\frac{1}{2}}\!{\hat\Phi}(x,y), 
\quad 
\Psi(x)=\left(\frac{M_*^2}{l_4/\delta}\right)^{\frac{1}{2}}\!{\hat\Psi}(x),
\quad
d^2\theta=M_* d^2{\hat\theta}, \ \ \theta=\frac{\hat\theta}{\sqrt{M_*}},
\eea
the brane action becomes

\be
{\cal L}_F=\frac{M_*^4}{l_4/\delta}\int d^2{\hat\theta}\, 
{\hat W}({\hat\Phi},{\hat\Psi})+{\rm h.c.}.\\
\nonumber
\ee
In particular, for  the superpotential for the Yukawa interaction 
\be
W=\lambda_t \,H_u\, Q \,U^c,
\qquad{\rm with} \qquad  \lambda_t=V^{p/2}_{D-4}\,f_{4,t}
\ee
where $p$ is the number of bulk fields present in the Yukawa
interaction, and $f_{4,t}$ is the 4D coupling.
Then, the redefined superpotential is given by
\be
{\hat W}=\lambda_{\rm eff} \,{\hat H}_u \,{\hat Q} \,{\hat U}^c
\ee
where 
\be
\lambda_{\rm eff}\equiv \lambda_t\, M_* \,\frac{l_4/\delta}{M_*^4}
\left(\frac{M_*^{D-2}}{l_D/\delta}\right)^{\frac{p}{2}} 
\left(\frac{M_*^2}{l_4/\delta}\right)^{\frac{3-p}{2}}\! <1
\ee
This condition gives the bound on $M_*$
\be
M_*<\left[\frac{(l_D/\delta)^p\, \,(l_4/\delta)^{1-p}}{f^2_{4,t}\, V_{D-4}^p}
\right]^{\frac{1}{p\,(D-4)}}
\label{ybound}
\ee
We  now apply eqs.(\ref{gbound}), (\ref{ybound}) 
to our 5D models, to derive bounds on the 5D cutoff $M_*$.
Eq.(\ref{gbound}) gives
\be\label{bound_g}
M_*<\frac{12\,\pi^2}{\delta \,g_4^2}\,\frac{1}{R}.
\ee 
Eq.~(\ref{ybound}) becomes for the 5D case
\bea\label{bound_y}
M_*< \frac{12\pi^2}{\delta}
\bigg[\frac{\delta}{16 \pi^2 \,f_{4,t}^{2/(p-1)}}\bigg]^{\frac{p-1}{p}}
\!\frac{1}{R}
\eea
where $p=2, 3$.
Eqs.(\ref{bound_g}), (\ref{bound_y}) were used in the text, 
eqs.(\ref{bnd}).

\end{document}